\documentstyle[aps,pra,preprint,tighten,eqsecnum]{revtex}
\input{psfig}

\begin{document}
\draft
\title{Semiclassical theory of shot noise in mesoscopic conductors}

\author{M. J. M. de Jong$^{a,b}$ and C. W. J. Beenakker$^{b}$}
\address{ 
(a) Philips Research Laboratories,
5656 AA  Eindhoven,
The Netherlands\\
(b) Instituut-Lorentz,
University of Leiden,
2300 RA  Leiden,
The Netherlands}
\date{\today}
\maketitle

\begin{abstract}
A semiclassical theory
is developed for time-dependent current fluctuations in mesoscopic conductors.
The theory is based on the Boltzmann-Langevin equation for a degenerate
electron gas.
The low-frequency shot-noise power is related to 
classical transmission probabilities at the Fermi level.
For a disordered conductor with impurity scattering,
it is shown how the shot noise crosses over from zero in the ballistic
regime to one-third of the Poisson noise in the diffusive regime.
In a conductor
consisting of $n$ tunnel barriers in series,
the shot noise approaches one-third
of the Poisson noise as $n$ goes to infinity,
independent of the transparency of the barriers.
The analysis confirms that phase coherence is not required for the
occurrence of the one-third suppression of the shot noise.
The effects of electron heating and inelastic scattering
are calculated, by inserting 
charge-conserving electron reservoirs between 
segments of the conductor.
\end{abstract}



\vfill

\newpage

\section{Introduction}
\label{s1}

The discreteness of the electron charge causes time-dependent
fluctuations in the electrical current, 
known as shot noise. These fluctuations are characterized by a white
noise spectrum and persist down to zero temperature.
The shot-noise power $P$ contains information on the conduction process,
which is not given by the resistance.
A well-known example is a vacuum diode, where
$P=2 e |\bar{I}|\equiv P_{{\rm Poisson}}$, with $\bar{I}$ the average current.
This tells us that the electrons traverse the conductor in a completely
uncorrelated fashion, as in a Poisson process.
In macroscopic samples, the shot noise is averaged out to zero by
inelastic scattering.

In the past few years, the shot noise has been investigated
in mesoscopic conductors, smaller than the
inelastic scattering length. Theoretical analysis
shows that the shot noise can be
suppressed below $P_{{\rm Poisson}}$, due to correlations in the electron
transmission imposed by the Pauli 
principle \cite{khl87,les89,yur90,but90,mar92}.
Most intriguingly, it has been found that 
$P=\case{1}{3} P_{{\rm Poisson}}$
in a metallic, diffusive
conductor \cite{b&b92,nag92,jon92,naz94,alt94}.
The factor one-third is universal in the sense that it is independent
of the material, sample size, or degree of disorder,
as long as the length $L$ of the conductor is greater than
the mean free path $\ell$ and shorter than the localization length.
An observation of
suppressed shot noise in a diffusive conductor has been 
reported \cite{lie94}.
In a quantum mechanical description \cite{b&b92},
the suppression follows from the
bimodal distribution of transmission eigenvalues \cite{dor84}.
Surprisingly, Nagaev \cite{nag92} finds the same one-third suppression
from a semiclassical approach, in which the Pauli principle is accounted
for, but 
the motion of electrons
is treated classically. 
This implies that
phase coherence is not essential for the suppression.
A similar conclusion is obtained in Ref.\ \cite{shi92}.
However, the relationship
between the quantum mechanical and
semiclassical theories remains unclear \cite{lan95b}.

In this paper, we reinvestigate the semiclassical
approach and present a detailed comparison with quantum mechanical
calculations in the literature.
In particular, we study how the shot noise crosses over from the ballistic
to the diffusive regime. This complements the study of the crossover of the
conductance in Ref.\ \cite{jon94}.
We use the
Boltzmann-Langevin equation \cite{kad57,kog69},
which is a semiclassical kinetic equation
for nonequilibrium fluctuations.
This equation has previously been applied to shot noise
by Kulik and Omel'yanchuk \cite{kul84} 
for a ballistic point contact,
and by Nagaev \cite{nag92} for a diffusive conductor.
Here, we will demonstrate how the
Boltzmann-Langevin equation can be applied to
an arbitrary mesoscopic conductor.
Our analysis corrects previous work by Beenakker and Van Houten
\cite{bee91}.
A brief account of our main results has
been reported in Ref.\ \cite{jon95}.

The outline of this paper is as follows:
In Section \ref{s2} we discuss the Boltzmann-Langevin equation.
It is demonstrated how the shot-noise power can be expressed in terms of 
semiclassical transmission probabilities.
Impurity scattering is treated in Section \ref{s3}. 
The shot-noise power increases from zero in the ballistic regime
to $\case{1}{3}  P_{\rm Poisson}$ in the diffusive regime.
We consider both 
isotropic and nonisotropic 
impurity scattering, and both a two- and
three-dimensional density of states.
We also present a one-dimensional model, which
can be solved analytically. 
Exact agreement is found with a previous
quantum mechanical evaluation \cite{jon92},
in the limit of a conductance $\gg e^2/h$.
Section \ref{s4} deals with barrier scattering.
We consider tunneling through $n$ planar barriers in series
(tunnel probability $\Gamma$).
For $n=2$ and $\Gamma \ll 1$, we recover the results for a 
double-barrier junction of Refs.\ \onlinecite{che91}
and \onlinecite{dav92}.
In the limit $n \rightarrow \infty$ the shot-noise power approaches
$\case{1}{3} P_{{\rm Poisson}}$ independent of $\Gamma$.
By taking the continuum limit, $n \rightarrow \infty$, 
$\Gamma \rightarrow 1$, at fixed $n (1-\Gamma)$,
we recover the one-dimensional model of Section \ref{s3}.
The case of a disordered region in series with a tunnel barrier concludes
Section \ref{s4}.
In Section \ref{s5} we calculate the effects of inelastic scattering and
of electron heating due to electron-electron scattering. Analogous to the
work of Beenakker and B\"{u}ttiker \cite{b&b92}, 
this scattering is modeled by putting
charge-conserving electron reservoirs between phase-coherent 
segments of the conductor. This allows us to model the effects of
quasi-elastic scattering, electron heating, and inelastic scattering within
a single theoretical framework.
We conclude in Section \ref{s6}.

Before proceeding with a description of the semiclassical approach,
we briefly summarize the fully quantum mechanical theory.
The zero-temperature, zero-frequency shot-noise power $P$ 
of a phase-coherent conductor is related to the
transmission matrix ${\sf t}$ by the formula \cite{but90}
\begin{equation}
P = P_0 \mbox{Tr} \,
 {\sf t} \, {\sf t}^\dagger
( {\sf 1} - {\sf t} \,
{\sf t}^\dagger )  
= P_0 \sum_{n=1}^{N}
T_n ( 1 - T_n ) \: ,
\label{e1}
\end{equation}
where $P_0\equiv 2 e |V| G_0$, 
with $V$ the applied voltage and $G_0 \equiv e^2/h$ the conductance
quantum (we assume spinless electrons for simplicity of notation),
$T_n \in [0,1]$ an eigenvalue of
${\sf t} \, {\sf t}^\dagger$, and $N$ the number of transverse
modes at the Fermi energy $E_F$. 
The conductance is given by the Landauer formula
\begin{equation}
G = G_0 \mbox{Tr} \, {\sf t} \, {\sf t}^\dagger
 = G_0 \sum_{n=1}^{N} T_n
\: .
\label{e2}
\end{equation}
If the conductor is such that all
$T_n \ll 1$ ({\em e.g.}, a high tunnel barrier),
one finds $P=2 e |V| G \equiv P_{{\rm Poisson}}$, corresponding to a Poisson
distribution of the emitted electrons. It has been demonstrated
by Levitov and Lesovik \cite{lev93} (see also Ref.\ \cite{lee95})
that the general formula (\ref{e1})
corresponds to a binomial (or Bernoulli) distribution of the emitted
electrons for each transmission eigenstate.
If some $T_n$ are near 1 (open channels), then the shot noise is
reduced below $P_{{\rm Poisson}}$.
This implies
that in a quantum point contact the shot noise is absent on the
plateaus of conductance quantization and appears only at the steps
between the plateaus \cite{les89}. This effect has indeed been observed in
experiments \cite{li90a,rez95,kum95}.
In a metallic, diffusive conductor the $T_n$ are either
exponentially small or of order unity \cite{dor84}.
This bimodal distribution is required by Ohm's law for the average
conductance \cite{mor} and has been derived microscopically by 
Nazarov \cite{naz94} and by Altshuler, Levitov, and Yakovets \cite{alt94}.
As a consequence of the bimodal distribution, the shot-noise power is
reduced to one-third of the Poisson noise \cite{b&b92}.

It has been emphasized by Landauer \cite{lan95a},
that Coulomb interactions may induce a further reduction
of $P$.
Here, we follow the quantum mechanical
treatments in assuming noninteracting electrons,
within the framework of the Boltzmann-Langevin approach. 
We do include the effects of 
electrostatic potential fluctuations in 
Section \ref{s5}.

\section{Boltzmann-Langevin equation}
\label{s2}

We begin by formulating the semiclassical kinetic theory \cite{kad57,kog69}.
We consider a  conductor with a $d$-dimensional density of states
connected by ideal leads to two electron reservoirs (see Fig.\ \ref{ff1}).
The reservoirs have a temperature $T_0$ and a voltage difference $V$.
The electrons in the left and the right reservoir are in  
equilibrium, with
distribution function,
$f_L(\varepsilon) = f_0 (\varepsilon - eV)$ and 
$f_R(\varepsilon) = f_0 (\varepsilon)$, respectively.
Here $f_0$ is 
the Fermi-Dirac distribution,
\begin{equation}
f_0(\varepsilon) = \left[ 1 + \exp \left( \frac{\varepsilon -
E_F}{T_0} \right) \right]^{-1} \: .
\label{2e1}
\end{equation}
The {\em fluctuating} distribution function
$f({\bf r}, {\bf k}, t)$ 
in the conductor
equals $(2 \pi)^d$ times the density
of electrons with position ${\bf r}$,
and wave vector ${\bf k}$, at time $t$.
[The factor $(2 \pi)^d$ is introduced so that $f$ is the occupation number
of a unit cell in phase space.]
The average over time-dependent fluctuations
$\langle f \rangle \equiv \bar{f}$
obeys the Boltzmann equation,
\begin{mathletters}
\label{e3}
\begin{equation}
\left( \frac{d}{dt} + {\cal S} \right) 
\bar{f}({\bf r}, {\bf k}, t) = 0 \: ,
\label{e3a}
\end{equation}
\begin{equation}
\frac{d}{dt} \equiv
\frac{\partial}{\partial t} + 
{\bf v} \cdot \frac{\partial}{\partial {\bf r}} + 
{\cal F} \cdot \frac{\partial}{\hbar \partial {\bf k}} \: .
\label{e3b}
\end{equation}
\end{mathletters}%
The derivative (\ref{e3b})
(with ${\bf v}=\hbar {\bf k}/m$)
describes the classical motion in the force field 
${\cal F}({\bf r})=-e \partial \phi({\bf r}) / \partial {\bf r}
+ e {\bf v} \times {\bf B}({\bf r})$,
with electrostatic potential $\phi({\bf r})$
and magnetic field ${\bf B}({\bf r})$.
The term ${\cal S} \bar{f}$ accounts for the
stochastic effects of scattering. 
Only elastic scattering is taken into account
and electron-electron scattering is disregarded.
In the case of impurity scattering,
the scattering term in the Boltzmann equation (\ref{e3})
is given by
\begin{eqnarray}
{\cal S} f({\bf r},{\bf k},t) &=&
\int d {\bf k}' \, W_{ {\bf k}{\bf k}'}({\bf r}) \left\{ 
f({\bf r},{\bf k},t) [ 1 - f({\bf r},{\bf k}',t) ] 
-
f({\bf r},{\bf k}',t)  [ 1 - f({\bf r},{\bf k},t) ] \right\}
\nonumber \\
&=& \int d {\bf k}' \, W_{ {\bf k}{\bf k}'}({\bf r}) [ 
f({\bf r},{\bf k},t)  - f({\bf r},{\bf k}',t) ]
\: .
\label{ee4}
\end{eqnarray}
Here, $W_{ {\bf k}{\bf k}'}({\bf r})$ is the transition rate 
for scattering from ${\bf k}$ to ${\bf k}'$, which may in
principle also depend on ${\bf r}$.
[We assume inversion symmetry, so that 
$W_{ {\bf k}{\bf k}'}({\bf r})=W_{ {\bf k}'{\bf k}}({\bf r})$.]

We consider the stationary situation, where
$\bar{f}$ is independent of $t$.
The time-dependent fluctuations
$\delta f \equiv f - \bar{f}$ satisfy
the Boltzmann-Langevin equation \cite{kad57,kog69},
\begin{equation}
\left( \frac{d}{dt} + {\cal S} \right)
\delta f ({\bf r}, {\bf k}, t) = 
j ({\bf r}, {\bf k}, t) \: ,
\label{e5}
\end{equation}
where $j$ is a fluctuating source term
representing the fluctuations induced by the stochastic nature of the
scattering.
The flux $j$ has zero average,
$\langle j \rangle = 0$, and covariance 
\begin{equation}
\langle j ({\bf r}, {\bf k}, t) \, j ({\bf r}', {\bf k}', t') \rangle
= 
(2 \pi)^d \, \delta({\bf r} -{\bf r'}) \, \delta(t-t')
J({\bf r}, {\bf k}, {\bf k}') \: .
\label{e6}
\end{equation}
The delta functions ensure that 
fluxes are only correlated
if they are induced by the same scattering process.
The flux correlator $J$ depends on the type of scattering and on $\bar{f}$, 
but not on $\delta f$.
The correlator $J$ for the 
impurity-scattering term (\ref{ee4}) has been
derived by Kogan and Shul'man \cite{kog69},
\begin{eqnarray}
J({\bf r}, {\bf k}, {\bf k'})&=&
- W_{ {\bf k}{\bf k}'}({\bf r}) \left[ \bar{f} ( 1-\bar{f}') + 
\bar{f}'(1-\bar{f}) \right]
\nonumber \\
&+&  \delta( {\bf k} - {\bf k}') \int d {\bf k}'' \,
W_{ {\bf k}{\bf k}''}({\bf r}) \left[ \bar{f} ( 1-\bar{f}'') + 
\bar{f}''(1-\bar{f}) \right]
\: ,
\label{ee5}
\end{eqnarray}
where $\bar{f}\equiv \bar{f}({\bf r},{\bf k})$,
$\bar{f}'\equiv \bar{f}({\bf r},{\bf k}')$, and
$\bar{f}''\equiv \bar{f}({\bf r},{\bf k}'')$.
One verifies that
\begin{equation}
\int \!  d {\bf k} \, J({\bf r},{\bf k},{\bf k}') = 
\int \!  d {\bf k}' \, J({\bf r},{\bf k},{\bf k}') = 0 \: ,
\label{ae14}
\end{equation}
as it should, since the fluctuating source term conserves the number of
particles
[$\int d{\bf k} \, j({\bf r}, {\bf k}, t) =0$].
For the derivation of Eq.\ (\ref{ee5}) we refer to Ref.\ \cite{kog69}.
In Section \ref{s4} we give a similar derivation for $J$ in the case of
barrier scattering.

Since $j$ and $j'$ are uncorrelated for $t > t'$, it follows from 
Eq.\ (\ref{e5}) that the correlation function
$\langle \delta f \, \delta f' \rangle$ satisfies a Boltzmann equation in
the variables ${\bf r}, {\bf k}, t$,
\begin{equation}
\left( \frac{d}{dt} + {\cal S} \right) 
\langle \delta f({\bf r}, {\bf k}, t)  \, \delta f({\bf r}', {\bf k}', t') 
\rangle = 0 \: .
\label{ee1}
\end{equation}
Equation (\ref{ee1}) forms the starting point of the method of
moments of Gantsevich, Gurevich, and Katilius \cite{gan79}.
This method is very convenient to study equilibrium fluctuations, 
because the equal-time correlation is known,
\begin{equation}
\langle \delta f({\bf r}, {\bf k}, t)
\delta f({\bf r}', {\bf k}', t) \rangle_{\rm equilibrium} 
= (2 \pi)^d \,
\bar{f}({\bf r}, {\bf k}) [ 1 - \bar{f}({\bf r}, {\bf k})]
\, \delta( {\bf k} - {\bf k'}) \, \delta( {\bf r} - {\bf r'} ) \: ,
\label{ee2}
\end{equation}
and Eq.\ (\ref{ee1}) can be used to compute the non-equal-time
correlation.
(For a study of thermal noise within this approach,
see, for example, Ref.\ \cite{bul93}.)
Out of equilibrium, Eq.\ (\ref{ee2}) does not hold, except in the
reservoirs, and one has to return to the full Boltzmann-Langevin equation
(\ref{e5}) to determine the shot noise.
In particular, it is only in equilibrium that the equal-time
correlation $\langle \delta f \, \delta f' \rangle$ vanishes
for ${\bf r} \neq {\bf r'}$, ${\bf k} \neq {\bf k'}$.
Out of equilibrium, scattering correlates fluctuations $\delta f$ at
different momenta and different points in space.

To obtain the shot-noise power we compute the current
$I(t) \equiv \bar{I} + \delta I(t)$ through a cross section
$S_R$  in the right lead.
The average current $\bar{I}$ and the fluctuations $\delta I(t)$
are given by
\begin{equation}
\bar{I} = \frac{e}{(2 \pi)^d} 
\int \limits_{S_R} \! \! d{\bf y} \int \! d {\bf k} \:
v_x \, \bar{f} ({\bf r}, {\bf k}) \: ,
\label{e7a}
\end{equation}
\begin{equation}
\delta I (t) = \frac{e}{(2 \pi)^d} 
\int \limits_{S_R} \! \! d{\bf y} \int \! d {\bf k} \:
v_x \, \delta f ({\bf r}, {\bf k}, t) \: .
\label{e7}
\end{equation}
We denote ${\bf r}=(x,{\bf y})$,
with the $x$-coordinate along and ${\bf y}$ 
perpendicular to the wire (see Fig.\ \ref{ff1}).
The zero-frequency noise power
is defined as
\begin{equation}
P \equiv 2 \int \limits_{-\infty}^{\infty} \! \!
dt \, \langle \delta I (t) \, \delta I (0) \rangle \: .
\label{e8}
\end{equation}
The formal solution of Eq.\
(\ref{e5}) is
\begin{equation}
\delta f ({\bf r}, {\bf k}, t)=
\int \limits_{-\infty}^{t} \! \! dt' \int \! d{\bf r}' \int \! d{\bf k}'
\, {\cal G}( {\bf r}, {\bf k}; {\bf r}', {\bf k}'; t-t') \,
j({\bf r}', {\bf k}', t') \: ,
\label{e9}
\end{equation}
where the Green's function $\cal G$ is a solution of
\begin{equation}
\left( \frac{d}{dt} + {\cal S} \right) 
{\cal G}( {\bf r}, {\bf k}; {\bf r}', {\bf k}';t)
= \delta({\bf r}-{\bf r}') \,
\delta({\bf k}-{\bf k}') \, \delta(t) \: ,
\label{e10}
\end{equation}
such that ${\cal G}=0$ if $t < 0$.
The transmission probability $T({\bf r}, {\bf k})$
is the probability that an electron at $({\bf r}, {\bf k})$
leaves the wire through the right lead.
It is related to ${\cal G}$ by
\begin{equation}
T({\bf r}, {\bf k}) = 
\int \limits_0^{\infty} \!\! dt 
\int \limits_{S_R} \!\! d{\bf y}' 
\int \! d{\bf k}' \, v_x' \,
{\cal G}( {\bf r}', {\bf k}'; {\bf r}, {\bf k};t) \: .
\label{e11}
\end{equation}
Substitution of Eqs.\ (\ref{e7}) and (\ref{e9}) into
Eq.\ (\ref{e8})
yield for the noise power the expression
\begin{eqnarray}
P&=&\frac{2 e^2}{ (2 \pi)^{2d}}
\int \limits_{S_R} \! \! d{\bf y} \int \! d {\bf k} \:
v_x \, 
\int \limits_{S_R} \! \! d{\bf y'} \int \! d {\bf k'} \:
v'_x \, 
\int \limits_{-\infty}^{t} \! \! dt'' \int \! d{\bf r}'' \int \! d{\bf k}''
\, {\cal G}( {\bf r}, {\bf k}; {\bf r}'', {\bf k}''; t-t'') \,
\nonumber \\
&\times&  
\int \limits_{-\infty}^{0} \! \! dt''' \int \! d{\bf r}''' \int \! 
d{\bf k}'''
\, {\cal G}( {\bf r'}, {\bf k'}; {\bf r}''', {\bf k}'''; -t''') \,
\langle j ({\bf r''}, {\bf k''}, t'') 
j ({\bf r}''', {\bf k}''', t''') \rangle \: ,
\label{e12a}
\end{eqnarray}
which can be simplified using Eqs.\ (\ref{e6}) and (\ref{e11}):
\begin{equation}
P=
\frac{2 e^2}{ (2 \pi)^d} \int \!  d{\bf r} \!  
\int \!  d {\bf k} \!  \int \!  d {\bf k}' \,
T({\bf r},{\bf k}) \, T({\bf r},{\bf k}')
J({\bf r},{\bf k},{\bf k}') \: .
\label{e12}
\end{equation}
Equation (\ref{e12}) applies generally to any conductor.
It contains the noise due to the current fluctuations induced by the
scattering processes inside the conductor. At non-zero temperatures,
there is an additional source of noise from fluctuations
which originate from the reservoirs.
In  Appendix \ref{a1} it is shown how this thermal noise can be 
incorporated.
In what follows, we restrict to zero temperature.

A final remark concerns the $x$-coordinate of the cross section at which
the current is evaluated [at $x=x_R$ in Eq.\ (\ref{e7})]. 
 From current conservation it follows that the zero-frequency
noise power should not depend on the specific value of $x$. 
This is explicitly proven in Appendix \ref{a2}, as a check on the
consistency of the formalism.

\section{Impurity scattering}
\label{s3}

In this Section we
specialize to elastic impurity scattering in a conductor made of a
material with a spherical Fermi surface
and in which the force field ${\cal F}=0$
(so we do not consider the case that a magnetic field is present). 
The conductor has a length $L$ and a
constant width $W$ ($d=2$) or a constant
cross-sectional area $A$ ($d=3$).
(In general expressions, both $W$ and $A$ will be denoted by $A$.)
We calculate the shot noise
at zero temperature and small applied voltage,
$e V \ll E_F$, so that we need to consider electrons at the
Fermi energy only.
The case of non-zero temperature is briefly discussed in Appendix
\ref{a1}.

It is useful to
change variables from wave vector
$\bf k$ to energy $\varepsilon= \hbar^2 k^2 /2 m$, 
and unit vector ${\bf \hat{n}}\equiv {\bf k}/k$.
The integrations are modified accordingly,
\begin{equation}
 \int \frac{d {\bf k}}{(2 \pi)^{d}}
\rightarrow
\int d\varepsilon \, {\cal D}(\varepsilon) 
\int \frac{d {\bf \hat{n}}}{s_d }
\: ,
\label{3ee1}
\end{equation}
where 
${\cal D}(\varepsilon)=s_d m (k/2\pi)^{d-2} h^{-2}$
is the density of states, and $s_d$ is
the surface of a $d$-dimensional unit
sphere ($s_1=2, \; s_2=2\pi, \; s_3=4 \pi$).
We consider the case of specular boundary scattering and 
assume that the elastic impurity-scattering rate
$W_{{\bf k}{\bf k}'}({\bf r})\equiv
W_{{\bf \hat{n}}{\bf \hat{n}}'}   
\delta(\varepsilon -\varepsilon')/ {\cal D}(\varepsilon)$
is independent of ${\bf r}$.
This allows us to drop the transverse coordinate $\bf y$ and write
$T( {\bf r}, {\bf k} ) = T( x, {\bf \hat{n}} )$ 
for the transmission probability at the Fermi level.
 From Eqs.\ (\ref{e10}) and (\ref{e11}) we derive a Boltzmann
equation for the transmission probability \cite{jon94},
\begin{mathletters}
\label{3e1}
\begin{eqnarray}
v_F \, n_x \, \frac{\partial T( x, {\bf \hat{n}} )}{\partial x}
&=& {\cal S} T( x, {\bf \hat{n}} ) \: 
\nonumber \\
&=& \int \frac{d {\bf \hat{n}}'}{s_d} \, W_{{\bf \hat{n}}{\bf \hat{n}}'}
 \,  [ T( x, {\bf \hat{n}} ) - T( x, {\bf \hat{n}}')] \: .
\label{3e1a}
\end{eqnarray}
The boundary conditions in the left and the right leads are
\begin{eqnarray}
T(0, {\bf \hat{n}}) &=& 0 \: , \; 
\mbox{ if } n_x < 0 \: ,
\\
T(L, {\bf \hat{n}}) &=& 1 \: , \; 
\mbox{ if } n_x > 0 \: ,
\end{eqnarray}
\end{mathletters}%
where $x_L=0$ and $x_R=L$ are the $x$-coordinates of the left and right
cross section $S_L$ and $S_R$, respectively.

The average distribution function can be expressed as
\begin{equation}
\bar{f}({\bf r}, {\bf k}) = 
[1-T({\bf r}, -{\bf k})] f_L(\varepsilon) +  
T({\bf r}, -{\bf k}) f_R(\varepsilon)
\: ,
\label{3e4}
\end{equation}
where (because of time-reversal symmetry in the absence of a magnetic
field)
$T({\bf r}, -{\bf k}) $ equals the probability that an electron
at $({\bf r}, {\bf k})$ has arrived there from the right reservoir.
Combining Eqs.\ (\ref{e7a}) and (\ref{3e4}), we obtain
the semiclassical Landauer formula for the linear-response conductance
$G \equiv \lim_{V \rightarrow 0} \bar{I}/V$ \cite{bee89},
\begin{eqnarray}
G&=& \frac{e^2}{(2 \pi)^d} 
\int \limits_{S_x} \! \! d{\bf y}
\int \! d {\bf k} \,
\left(\! - \frac{\partial f_0(\varepsilon)}{\partial \varepsilon} \right)
v_x \, 
T({\bf r}, {\bf k}) 
\nonumber \\
&=&
e^2 \int \limits_{S_x} \! \! d{\bf y} 
\int d\varepsilon \, {\cal D}(\varepsilon)
\left(\! - \frac{\partial f_0(\varepsilon)}{\partial \varepsilon} \right)
\int \frac{d {\bf \hat{n}}}{s_d} \, v  \, n_x \, 
T(x, {\bf \hat{n}}) 
\nonumber \\
&=&
G_0 N  
\int  \frac{d {\bf \hat{n}}}{v_{d-1}} \, n_x \, 
T(x, {\bf \hat{n}}) \: ,
\label{3e5}
\end{eqnarray}
with $S_x$ the cross section at $x$.
The number of transverse modes 
$N=v_{d-1}(k_F/2\pi)^{d-1} A$, where $v_d$ is the volume of a $d$-dimensional
unit sphere ($v_0=1, \; v_1=2, \; v_2 = \pi$).
One has $N=k_F W /\pi$ for $d=2$ and
$N=k_F^2 A/ 4 \pi$ for $d=3$.
One can verify that 
the conductance in
Eq.\ ({\ref{3e5}) is independent of the
value of $x$, as it should:
By integrating Eq.\ (\ref{3e1a}) 
over $\bf \hat{n}$
one finds that 
\begin{equation}
\frac{\partial}{\partial x} \int  d {\bf \hat{n}} \, n_x \, 
T(x, {\bf \hat{n}}) = 0 \: .
\label{3e6}
\end{equation}

We evaluate the noise power by substitution of Eqs.\
(\ref{ee5}) and (\ref{3e4}) into Eq.\ (\ref{e12}).
Some intermediate steps are given in Appendix \ref{a1}.
The resulting zero-temperature shot-noise power is
\begin{equation}
P = \frac{P_0 N}{v_F s_d v_{d-1}}
\int \limits_0^L \! \! dx
\int  \! \! d {\bf \hat{n}} 
\int  \! \! d {\bf \hat{n}}' \,
W_{{\bf \hat{n}}{\bf \hat{n}}'} \,
[ T(x,{\bf \hat{n}}) - T(x,{\bf \hat{n}}')]^2  
T(x,-{\bf \hat{n}})[1-T(x,-{\bf \hat{n}}')]
\: .
\label{3e7}
\end{equation}
This completes our general semiclassical theory.
What remains is to compute 
the transmission probabilities from Eqs.\ (\ref{3e1})
for a particular choice of the scattering rate $W$. 
Comparing Eqs.\ (\ref{e2}) and (\ref{3e5}), we note
that $\sum_n T_n$ corresponds semiclassically
to $N \int d{\bf \hat{n}} \, n_x T(x,{\bf \hat{n}})$.
Comparison of Eqs.\ (\ref{e1}) and (\ref{3e7}) 
shows that the semiclassical
correspondence to $\sum_n T_n(1-T_n)$ is much more complicated, as it
involves the transmission probabilities $T(x,{\bf \hat{n}})$ 
at all scatterers inside
the conductor (and not just the transmission probability 
$T(0,{\bf \hat{n}})$ through
the whole conductor).

In a ballistic conductor, where impurity scattering is absent,
the transmission probabilities are given by
$T(x,{\bf \hat{n}})=1$, if $n_x>0$, and $T(x,{\bf \hat{n}})=0$, if $n_x<0$.
 From Eq.\ (\ref{3e5}), we then obtain the 
Sharvin conductance $G_S\equiv G_0 N$ \cite{sha65}.
Equation (\ref{3e7}) implies that the shot-noise power is zero, in
agreement with a previous semiclassical calculation by
Kulik and Omel'yanchuk \cite{kul84}. 

We now restrict ourselves to the case $W_{{\bf \hat{n}}{\bf \hat{n}}'}
=v_F / \ell$ of isotropic impurity scattering.
Let us first show that in the diffusive limit ($\ell \ll L$) the result of
Nagaev \cite{nag92} is recovered.
For a diffusive wire the solution of Eq.\ (\ref{3e5}) can be approximated
by
\begin{equation}
T(x,{\bf \hat{n}})= \frac{ x + \ell \, n_x}{L} \: .
\label{3e8}
\end{equation}
Deviations from
this approximation only occur within a thin layer, of order $\ell$,
at the ends $x=0$ and $x=L$. 
Substitution of  Eq.\ (\ref{3e8})
into Eq.\ (\ref{3e5}) yields the Drude conductance
\begin{equation}
G_D= G_0 N \, \frac{\tilde{\ell}}{L} \: ,
\label{3e8a}
\end{equation}
with the normalized mean free path 
$\tilde{\ell} \equiv (v_d /  v_{d-1}) \ell$, 
i.e.\ for $d=2$ we have $\tilde{\ell}=\case{1}{2} \pi \ell$ and for
$d=3$ we have $\tilde{\ell}=\case{4}{3} \ell$.
For the shot-noise power we obtain from Eq.\ (\ref{3e7}), neglecting
terms of order $(\ell/L)^2$,
\begin{equation}
P= P_0 N \frac{2 \tilde{\ell}}{L} \int \limits_0^L \frac{dx}{L} 
\, \frac{x}{L}(1- \frac{x}{L}) = \case{1}{3} P_{\rm Poisson}
\: ,
\label{3e9}
\end{equation}
in agreement with Nagaev \cite{nag92}.
This result is a direct consequence of the linear dependence of the
transmission probability (\ref{3e8}) on $x$, which is generic for
diffusive transport.
In Appendix \ref{a3} it is demonstrated
that for a diffusive conductor with arbitrary 
(nonisotropic) impurity scattering 
$W_{{\bf \hat{n}}{\bf \hat{n}}'}$, the result $P=\case{1}{3} P_{\rm Poisson}$
remains valid.

We can go beyond Ref.\ \cite{nag92} and apply our method to
quasi-ballistic conductors, for which $\ell$ and $L$ become comparable.
In Ref.\ \cite{jon94}, we showed how in this case the probability 
$T(x,{\bf \hat{n}})$ can be calculated numerically by solving Eq.\ 
(\ref{3e1}).
With this numerical solution as input, we compute the conductance and the
shot-noise power from Eqs.\ (\ref{3e5}) and (\ref{3e7}).
The result is shown in Fig.\ \ref{ff2}.
The conductance crosses over from the 
Sharvin conductance to the Drude conductance with increasing length
\cite{jon94}. This crossover is accompanied by a rise in the shot noise,
from zero to $\case{1}{3} P_{\rm Poisson}$.
We note small differences between the two and the three-dimensional case
in the crossover regime.
The crossover is only weakly dependent on the dimensionality of the Fermi
surface.

The dimensions $d=2$ and 3 require a numerical solution of 
Eqs.\ (\ref{3e1}). For $d=1$ an analytical solution is possible.
We emphasize that this is not a model for true one-dimensional
transport, where quantum interference leads to
localization if $L > \ell$ \cite{mel95}.
The case $d=1$ should rather be considered as
a toy model, which displays similar 
behavior as the two and three-dimensional cases,
but which allows us to
evaluate both the conductance and the shot-noise power analytically for
arbitrary ratio $\ell / L$.
In the case $d=1$ an electron
can move either forward or backward, so
$n_x \equiv n$ is either 1 or $-1$.
The solution of Eq.\ (\ref{3e1}) is
\begin{equation}
T(x,n)=\frac{x + \ell (n+1)}{L+2\ell} \: .
\label{3e10}
\end{equation}
Substitution into Eq.\ (\ref{3e5}) yields
\begin{equation}
G= G_0 N \, \frac{1}{1 + L / \tilde{\ell}}
\: ,
\label{3e11}
\end{equation}
where $\tilde{\ell} \equiv 2 \ell$. 
Note that the resistance $1/G$ is precisely the sum of the
Drude and the Sharvin resistance.
The shot-noise power follows from Eq.\ (\ref{3e7}),
\begin{eqnarray}
P &=& P_0 N \, 
\frac{\case{2}{3} \ell L^3 + 4 \ell^2 L^2+  8 \ell^3 L}
{( L + 2 \ell )^4}
\nonumber \\
&=&\case{1}{3} 
\left( 1 -
\frac{1}{(1 + L/\tilde{\ell})^{3}} \right)  P_{{\rm Poisson}}
\: .
\label{3e12}
\end{eqnarray}
In Fig.\ \ref{ff2} we have plotted $G$ and $P$
according to Eqs.\ (\ref{3e11}) 
and (\ref{3e12}). 
The difference with $d=2$ and $d=3$ is very small.

Liu, Eastman, and Yamamoto \cite{liu95b} 
have carried out Monte Carlo simulations of the
shot noise in a mesoscopic conductor, in good agreement with
Eq.\ (\ref{3e12}).
In Ref.\ \cite{jon92}, we have performed a quantum mechanical study of
the shot noise in a wire geometry, on the basis
of the Dorokhov-Mello-Pereyra-Kumar equation \cite{dor82}.
The semiclassical results for $d=1$ obtained in the present paper, both for
the conductance and for the shot-noise power, coincide precisely with these
quantum mechanical results, in the limit $N \tilde{\ell}/L \gg 1$.
Corrections (of order $P_0$) to the shot-noise power,
due to weak localization \cite{jon92}, are beyond the
semiclassical approach.

\section{Barrier scattering}
\label{s4}

We now specialize to the case 
that the scattering is due to $n$
planar tunnel
barriers in series, perpendicular to the $x$-direction 
(see inset of Fig.\ \ref{f1}).
Barrier $i$ has tunnel probability $\Gamma_i \in [0,1]$, 
which for simplicity
is assumed to be $\bf k$ and ${\bf y}$-independent.
In what follows, we again drop the $\bf y$-coordinate.
Upon transmission $\bf k$ is conserved, whereas upon reflection
${\bf k} \rightarrow \widetilde{\bf k}\equiv(-k_x,{\bf k}_y)$.
At barrier $i$ (at $x=x_i$)
the average densities $\bar{f}$ 
on the left side ($x_{i-}$) and on the
right side ($x_{i+}$) are related by
\begin{mathletters}
\label{e13}
\begin{eqnarray}
\bar{f}(x_{i+},{\bf k}) &=& \Gamma_i \bar{f}(x_{i-},{\bf k})
+ (1 - \Gamma_i)  \bar{f}(x_{i+},\widetilde{\bf k}) \: , \:
\mbox{ if } \: k_x > 0 \: ,
\\
\bar{f}(x_{i-},{\bf k}) &=& \Gamma_i \bar{f}(x_{i+},{\bf k})
+ (1 - \Gamma_i)  \bar{f}(x_{i-},\widetilde{\bf k}) \: , \: 
\mbox{ if } \: k_x < 0 \: .
\end{eqnarray}
\end{mathletters}%
To determine the correlator $J$ in Eq.\ (\ref{e6}), we argue in a similar
way as in Ref.\ \onlinecite{mar92}.
Consider an incoming state from the left $(x_{i-},{\bf k})$
and from the right $(x_{i+},\widetilde{\bf k})$
(we assume $k_x > 0$). 
We need to distinguish between four different situations:
\begin{enumerate}
\item[(a)] 
Both incoming states empty, probability
$[1-\bar{f}(x_{i-},{\bf k})] [ 1 - \bar{f}(x_{i+},\widetilde{\bf k}) ]$.
Since no fluctuations in the outgoing states are possible,
the contribution to $J$ is zero.
\item[(b)] 
Both incoming states occupied, probability
$\bar{f}(x_{i-},{\bf k}) \, \bar{f}(x_{i+},\widetilde{\bf k})$.
Again, no contribution to $J$.
\item[(c)]
Incoming state from the left occupied and from the right empty,
probability 
$\bar{f}(x_{i-},{\bf k}) [ 1 - \bar{f}(x_{i+},\widetilde{\bf k}) ]$.
On the average, the outgoing states at the left and right 
have occupation $1-\Gamma_i$ and $\Gamma_i$, respectively. 
However, since the incoming electron is either 
transmitted or reflected, the instantaneous occupation of the
outgoing states differs from the average occupation. Upon transmission,
the state at the right (left) has an excess (deficit) occupation of 
$1-\Gamma_i$. Upon reflection, 
the state at the right (left) has an deficit (excess) occupation of 
$\Gamma_i$.
Since transmission occurs with probability $\Gamma_i$ and reflection 
with  probability $1-\Gamma_i$,
the equal-time
correlation of the occupations is given by
\begin{equation}
\langle \delta f({\bf r}, {\bf k}, t)  \, \delta f({\bf r}', {\bf k}', t) 
\rangle = 
\left\{
\begin{array}{l}
(2 \pi)^d \,
\Gamma_i (1-\Gamma_i) \, \delta( {\bf k} - {\bf k'} ) \,
\delta( {\bf r} - {\bf r}' ) 
\: , \\
\;\;\; 
\vphantom{\int \limits_0^L}
\mbox{ if }  \;\; x,x' > x_i \: \mbox{ or } \: x, x' < x_i \: ,
\\
- (2 \pi)^d  \, \Gamma_i (1-\Gamma_i) \, 
\delta( {\bf k} - \widetilde{\bf k}' ) 
\delta( {\bf y} - {\bf y}' ) \, \delta( x + x' - 2 x_i ) 
\: , 
\\
\;\;\; \vphantom{\int \limits^L}
\mbox{ if } \;\; x < x_i < x' \: \mbox{ or } \:  
 x' < x_i < x  \: ,
\end{array} \right. 
\label{4e1}
\end{equation}
In terms of the fluctuating source,
the fluctuating occupation number can be expressed as
\begin{equation}
\delta f({\bf r}, {\bf k}, t) = \frac{1}{| v_x|}
\int dx_0 \, j \!\! \left( x_0, {\bf y} - \frac{ {\bf v}_y}{v_x} \, (x-x_0),
{\bf k}, t - \frac{x-x_0}{v_x} \right) \: ,
\label{4e2}
\end{equation}
where we have used Eq.\ (\ref{e9}).
(This result is valid as long as only one scattering event has occurred.)
Combining Eqs.\ (\ref{4e1}) and (\ref{4e2}),
it is found that 
\begin{eqnarray}
\langle j({\bf r}, {\bf k}, t)  \, j({\bf r}', {\bf k}', t') 
\rangle &=&
(2 \pi)^d \,
\Gamma_i (1-\Gamma_i) \,
\delta(x-x_i) \, |v_x| \,
\delta( {\bf r} - {\bf r}' ) 
\nonumber \\ &\times&
[ \delta( {\bf k} - {\bf k'} ) - \delta( {\bf k} - \widetilde{\bf k}' )]
\, \delta( t-t') \: ,
\label{4e3}
\end{eqnarray}
upon the initial condition of occupied left and
unoccupied right incoming state.
\item[(d)] 
Incoming state from the left unoccupied and from the right occupied,
probability 
$[1- \bar{f}(x_{i-},{\bf k})] \bar{f}(x_{i+},\widetilde{\bf k})$.
Similar to situation (c).
\end{enumerate}
Collecting results from (a)--(d) and summing over all barriers, we find
\begin{mathletters}
\label{e14}
\begin{eqnarray}
J(x,{\bf k},{\bf k}')&=& \sum_{i=1}^n
\delta(x-x_i) \, \Gamma_i (1-\Gamma_i) \, |v_x| \,
[ \delta({\bf k}-{\bf k}') - \delta({\bf k}-\widetilde{\bf k}')  ]
\nonumber \\
&\times&
\Bigl\{ \bar{f}(x_{i-},{\bf k}) \, [\, 1 - \bar{f}(x_{i+},\widetilde{\bf k})
 \, ] + 
\bar{f}(x_{i+},\widetilde{\bf k}) \, [\, 1 - \bar{f}(x_{i-},{\bf k}) \, ] 
\Bigr\}
\: ,
\nonumber \\ \mbox{if}&&
k_x > 0 \: ,
\\
J(x,{\bf k},{\bf k}')&=& \sum_{i=1}^n
\delta(x-x_i) \, \Gamma_i (1-\Gamma_i) \, |v_x| \,
[ \delta({\bf k}-{\bf k}') - \delta({\bf k}-\widetilde{\bf k}')  ]
\nonumber \\
&\times&
\Bigl\{ \bar{f}(x_{i+},{\bf k}) \, [\, 1 - \bar{f}(x_{i-},\widetilde{\bf k})
 \, ] + 
\bar{f}(x_{i-},\widetilde{\bf k}) \, [\, 1 - \bar{f}(x_{i+},{\bf k}) \, ] 
\Bigr\}
\: , 
\nonumber \\ \mbox{if}&& 
k_x < 0 \: .
\end{eqnarray}
\end{mathletters}%
Substitution of Eqs.\ (\ref{3e4}) and (\ref{e14})
into Eq.\ (\ref{e12}) and linearization in $V$ yields
\begin{equation}
P = P_0 N \,
\sum_{i=1}^n \Gamma_i (1-\Gamma_i) \,
(T_i^\rightarrow-T_i^\leftarrow)^2 (T_i^\rightarrow + 
T_i^\leftarrow - 2 T_i^\rightarrow T_i^\leftarrow )
\: ,
\label{e16}
\end{equation}
where $T_i^\rightarrow\equiv T(x_{i+},k_x>0)$
[$T_i^\leftarrow\equiv T(x_{i-},k_x<0)$] is the transmission probability
into the right reservoir
of an electron at the Fermi level
moving away from the right [left] side of barrier $i$.  
The conductance is given simply by
\begin{equation}
G=G_0 N \, T \: ,
\label{e17}
\end{equation}
where $T\equiv T(x_{1-},k_x>0)$ is the 
transmission probability through the whole conductor.

As a first application of Eq.\ (\ref{e16}), we calculate the shot noise for
a single tunnel barrier. Using $T=\Gamma$, 
$T_1^\leftarrow=0$, $T_1^\rightarrow=1$, we find 
the expected result \cite{khl87,les89,yur90,but90,mar92} 
$P=P_0 N \Gamma(1-\Gamma) = (1-\Gamma)
P_{{\rm Poisson}}$. 
The double-barrier case ($n=2$) is less trivial.
Experiments by Li {\em et al}.\ \cite{li90} 
and by Liu {\em et al}.\ \cite{liu95}
showed full Poisson noise,
for asymmetric structures ($\Gamma_1 \ll \Gamma_2$)
and a suppression by
one half, for the symmetric case ($\Gamma_1 \simeq \Gamma_2$).
This effect has been explained by 
Chen and Ting \cite{che91}, by Davies {\em et al}.\ \cite{dav92}, 
and by others \cite{her92}. 
These theories assume resonant tunneling in the regime that the applied
voltage $V$ is much greater than the width of the resonance.
This requires 
$\Gamma_1, \Gamma_2 \ll 1$. The present semiclassical 
approach makes no reference to transmission resonances and
is valid for all
$\Gamma_1, \Gamma_2$.
For the double-barrier system one has
$T=\Gamma_1 \Gamma_2/\Delta$,
$T_1^\leftarrow=0$,
$T_1^\rightarrow=\Gamma_2/\Delta$,
$T_2^\leftarrow=(1-\Gamma_1)\Gamma_2/\Delta$, and
$T_2^\rightarrow=1$, with $\Delta=\Gamma_1+\Gamma_2 - \Gamma_1 \Gamma_2$.
 From Eqs.\ (\ref{e16}) and (\ref{e17}), it follows that
\begin{equation}
P = 
\frac{\Gamma_1^2 (1-\Gamma_2) + \Gamma_2^2 (1-\Gamma_1)}
{(\Gamma_1+\Gamma_2 - \Gamma_1 \Gamma_2)^2} \, P_{{\rm Poisson}} 
\: .
\label{e19}
\end{equation}
In the limit $\Gamma_1, \Gamma_2 \ll 1$,
Eq.\ (\ref{e19}) coincides precisely with the results of Refs.\
\onlinecite{che91} and \onlinecite{dav92}.

The shot-noise suppression of one half for a symmetric
double-barrier junction has the same origin
as the one-third suppression for a diffusive conductor.
In our semiclassical model, this is evident from the
fact that a diffusive conductor is the continuum limit
of a series of tunnel barriers. We demonstrate this below.
Quantum mechanically, the common origin is the bimodal distribution
$\rho(T)\equiv \langle \sum_{n} \delta(T-T_n) \rangle$
of transmission eigenvalues, which for a double-barrier junction is
given by \cite{mel94}
\begin{equation}
\rho(T)= 
\frac{N \Gamma_1 \Gamma_2}{\pi T 
\sqrt{4 \Gamma_1 \Gamma_2 \, T - (\Delta \, T + \Gamma_1 \Gamma_2)^2}}
\: ,
\label{e20}
\end{equation}
for $T \in [T_-, T_+]$, with
$T_\pm=\Gamma_1 \Gamma_2 / (1 \mp \sqrt{1-\Delta} )^2$.
For a symmetric junction ($\Gamma_1=\Gamma_2\ll 1$), 
the density (\ref{e20}) is
strongly peaked near $T=0$ and $T=1$, leading to a suppression of shot noise,
just as in the case of a diffusive conductor. In fact, one can verify that
the average of Eqs.\ (\ref{e1}) and (\ref{e2}),
with the bimodal distribution (\ref{e20}), 
gives precisely the result (\ref{e19}) from the Boltzmann-Langevin equation.

We now consider $n$ barriers 
with equal $\Gamma$.
We find
$T=\Gamma/\Delta$,
$T_i^\rightarrow=[\Gamma+ i(1-\Gamma)]/\Delta$,
and 
$T_i^\leftarrow=(i-1)(1-\Gamma)/\Delta$, with
$\Delta=\Gamma+ n(1-\Gamma)$.
Substitution  into Eqs.\ (\ref{e16}) and (\ref{e17}) yields 
\begin{equation}
P = \frac{1}{3} 
\left( 1 +
\frac{n (1-\Gamma)^2 (2+\Gamma) - \Gamma^3}
{[\Gamma + n (1-\Gamma)]^3} \right) P_{{\rm Poisson}}
\: .
\label{e22}
\end{equation}
The shot-noise suppression for a low barrier ($\Gamma=0.9$)
and for a high barrier ($\Gamma=0.1$) is plotted 
against $n$ in Fig.\ \ref{f1}.
For $\Gamma=0.1$ we observe almost full shot noise if $n=1$,
one-half suppression if $n=2$, and on increasing $n$
the suppression rapidly reaches one-third.
For $\Gamma=0.9$, we observe that $P/P_{{\rm Poisson}}$ increases
from almost zero to one-third.
It is clear from Eq.\ (\ref{e22}) that  
$P \rightarrow \case{1}{3} P_{{\rm Poisson}}$
for $n \rightarrow \infty$
{\em independent} of $\Gamma$.

We can make the connection with elastic impurity scattering
in a disordered wire as follows:
The scattering occurs throughout the whole wire instead of
at a discrete number of barriers.
For the semiclassical evaluation
we thus take the limit $n \rightarrow \infty$ and
$\Gamma \rightarrow 1$, such that
$n (1-\Gamma) = L/\tilde{\ell}$.
For the conductance and the shot-noise power one then obtains
from Eqs.\ (\ref{e17}) and (\ref{e22}) exactly the  
same results, Eqs.\ (\ref{3e11}) and (\ref{3e12}), 
as for impurity scattering with a one-dimensional 
density of states.
This equivalence is expected, since in the one-dimensional 
model electrons  move either forward or backward,
whereas in the model of $n$ planar tunnel barriers in
series the transverse component of the wave vector becomes irrelevant.

We conclude this Section by considering a wire consisting of a disordered
region, between $x=0$ and $x=L$ with mean free path $\ell$, 
in series with a barrier, at $x=x_b > L$ with transparency $\Gamma$.
For analytical convenience, we study the one-dimensional case $d=1$.
(We have seen earlier that the dependence on $d$ is quite weak.)
By modifying Eqs.\ (\ref{3e1})
and (\ref{e13}), we find
\begin{mathletters}
\label{4e10}
\begin{eqnarray}
T(x,n) &=& \frac{x + \ell (1+n)}{L \Gamma + 2 \ell}\, \Gamma \: ,
\hspace{1cm} \mbox{if } \; x \in [0,L] \: ,
\\
T(x,n) &=& T(x,L) \: ,
\hspace{2.3cm} \mbox{if } \; x \in [L,x_b) \: ,
\\
T(x,-1) &=&  1 - \frac{2 \ell \Gamma}{L \Gamma + 2 \ell}\: ,
\hspace{1.4cm} \mbox{if } \; x > x_b \: ,
\\
T(x,1) &=& 1\: .
\hspace{3.5cm} \mbox{if } \; x > x_b \: ,
\end{eqnarray}
\end{mathletters}%
The conductance is given by Eq.\ (\ref{3e5}),
\begin{equation}
G= G_0 N \frac{\Gamma}{1 + \Gamma L/\tilde{\ell}} \: .
\label{4e11}
\end{equation}
The total resistance is thus the sum of the Drude resistance 
$R_D = L/G_0N\tilde{\ell}$  and the barrier
resistance $R_\Gamma = 1/G_0 N \Gamma$.
Combining Eqs.\ (\ref{3e7}) and (\ref{e16}), we obtain for the 
shot-noise power
\begin{eqnarray}
P &=& \frac{P_0 N}{2 \ell} \int \limits_0^L dx 
[ T(x,1)-T(x,-1) ]^2
\nonumber \\ &&\mbox{} \times
[ T(x,1) + T(x,-1) - 2 T(x,1) T(x,-1) ] 
\nonumber \\
\vphantom{\int \limits_0^L}
&+& P_0 N \, \Gamma (1-\Gamma) \, [ T(x_{b+},1)-T(x_{b-},-1) ]^2
\nonumber \\ &&\mbox{}  \times
[ T(x_{b+},1) + T(x_{b-},-1) - 2 T(x_{b+},1) T(x_{b-},-1) ] 
\: .
\label{4e12}
\end{eqnarray}
Substitution of Eqs.\ (\ref{4e10}) yields
\begin{eqnarray}
P&=& P_0 N \left( 
\frac{2 \Gamma^3 \ell L (12 \ell^2 + 6 \ell L + \Gamma L^2)}
{ 3 (2 \ell + \Gamma L)^4} +
\frac{8 \Gamma(1-\Gamma) \ell^3}{ (2 \ell + \Gamma L)^3}
\right)
\nonumber \\
&=& \left( \frac{1}{3} - \frac{1}{3(1+\Gamma L / \tilde{\ell})^3} +
\frac{1-\Gamma}{(1+\Gamma L / \tilde{\ell})^3 } \right) P_{\rm Poisson} 
\: ,
\label{4e13}
\end{eqnarray}
where we have used Eq.\ (\ref{4e11}).
In Fig.\ \ref{ff3} we have plotted the shot-noise power
against the length of the disordered region
for various values of the barrier transparency.
In the absence of disorder, there is full shot noise for
high barriers ($\Gamma \ll 1$)
and complete suppression if the barrier is
absent ($\Gamma=1$). 
Upon increasing the disorder strength, we note that the shot-noise power
approaches the limiting value $P=\case{1}{3} P_{\rm Poisson}$ 
{\em independent} of $\Gamma$:
Once the disordered region dominates the resistance, the shot noise
is suppressed by one-third.
Note, that it follows from Eq.\ (\ref{4e13}) that for $\Gamma=\case{2}{3}$
the suppression is one-third for all ratios $\tilde{\ell}/L$.

We have carried out a quantum mechanical calculation of the shot-noise
power in a wire geometry
similar to the calculation in Ref.\  \cite{jon92}.
The barrier can be incorporated in the 
Dorokhov-Mello-Pereyra-Kumar equation \cite{dor82}
by means of an initial condition (see Ref.\ \cite{bee94b}).
We find
exactly the same result as Eq.\ (\ref{4e13}) in the regime
$N\Gamma\gg1$ and $N\tilde{\ell}/L \gg 1$. 
For a high barrier ($\Gamma \ll 1$) in series with a diffusive wire 
($L \gg \tilde{\ell}$) our results for the shot noise
coincide with previous work by Nazarov \cite{naz94}
using a different quantum mechanical theory.
In this limit, the shot noise can be expressed as \cite{naz94}
\begin{equation}
P = \case{1}{3} \left[ 1 + 2 \left( \frac{R_\Gamma}{R} \right)^3 \right] 
P_{\rm Poisson} \: ,
\label{4e14}
\end{equation}
with the total resistance $R= R_D + R_\Gamma$.
The limiting result (\ref{4e14}) is depicted by the
dashed curve in Fig.\ \ref{ff3}.

\section{Inelastic and electron-electron scattering}
\label{s5}

In the previous Sections we have calculated the shot noise for several
types of elastic scattering.
In an experiment, however, additional types of scattering may occur.
In particular, electron-electron and inelastic 
electron-phonon scattering
will be enhanced due to the high currents 
which are often required for noise experiments.
The purpose of this Section is to discuss the effects of
these additional scattering processes.
As shown by Nagaev \cite{nag95} and by Kozub and Rudin \cite{koz95},
this can be achieved by including additional
scattering terms in the Boltzmann-Langevin equation.
Here, we will adopt a different method, following
Beenakker and B\"{u}ttiker \cite{b&b92},
in which inelastic scattering is modeled by dividing the conductor in
separate, phase-coherent
parts which are connected by charge-conserving reservoirs.
We extend this model to include
the following types of scattering:
\begin{itemize}
\item[(a)]
{\em Quasi-elastic scattering.} Due to weak coupling with external
degrees of freedom
the electron wave function gets dephased, but its energy is conserved. 
In metals, this scattering is caused by
fluctuations in the electromagnetic field  \cite{emf}.
\item[(b)]
{\em Electron heating.}
Electron-electron scattering ex\-changes energy
between the electrons, but the total energy of the electron system is conserved.
The distribution function is therefore assumed to be a 
Fermi-Dirac distribution at a temperature above the lattice temperature.
\item[(c)]
{\em Inelastic scattering.} Due to electron-phonon interactions the
electrons exchange energy with the lattice. The electrons emerging from
the reservoir are distributed according to the Fermi-Dirac distribution
(\ref{2e1}), at the lattice temperature $T_0$.
This is the model of Ref.\ \cite{b&b92}.
\end{itemize}
First, we divide the conductor in two parts connected via
one reservoir and determine the shot noise for case (a), (b), and (c).
After that, we repeat the calculation for many intermediate
reservoirs to
take into account that the scattering occurs throughout the whole
length of the conductor.

The model is depicted in Fig.\ \ref{s5f1}. The
conductors 1 and 2 are connected via a reservoir
with distribution function $f_{12}(\varepsilon)$.
The time-averaged current $I_m$ through conductor $m=1,2$ is given by 
\begin{mathletters}
\label{s5e1}
\begin{eqnarray}
I_1 &=& (G_1/e) \int \!\! d\varepsilon \, 
[ f_L(\varepsilon) - f_{12}(\varepsilon)]
\: , 
\label{s5e1a} \\
I_2 &=& (G_2/e) \int \!\! d\varepsilon \, 
[f_{12}(\varepsilon) - f_{R}(\varepsilon)]
\: .
\label{s5e1b}
\end{eqnarray}
\end{mathletters}%
The conductance $G_m\equiv1/R_m$ 
is expressed in terms of the
transmission matrix ${\sf t}_m$ of conductor $m$ at the Fermi energy,
\begin{equation}
G_m =G_0 \mbox{Tr} \, {\sf t}_m^{\vphantom{\dagger}} {\sf t}_m^\dagger = 
G_0 \sum_{n=1}^N T_n^{(m)} \: ,
\label{s5e2}
\end{equation}
with $T_n^{(m)} \in [0,1]$ an eigenvalue of 
${\sf t}_m^{\vphantom{\dagger}} {\sf t}_m^\dagger$.
We assume small $e V$ and $k_B T_0$, so that we can neglect 
the energy dependence of the transmission eigenvalues.

Current conservation requires that 
\begin{equation}
I_1=I_2\equiv I \: .
\label{s5e3}
\end{equation}
We define the total resistance of the conductor by
\begin{equation}
R=R_1+R_2 \: .
\label{s5e4}
\end{equation}
It will be shown that
this incoherent addition of resistances is valid for all three types
of scattering that we consider.
Our model is not suitable for
transport in the ballistic regime or in the quantum Hall regime,
where a different type of ``one-way'' reservoirs
are required \cite{but88}.
Recently, B\"{u}ttiker has calculated the effects of inelastic scattering
along these lines \cite{but95}.

The time-averaged current (\ref{s5e1}) depends on the
average distribution $f_{12}(\varepsilon)$ in the reservoir
between conductors 1 and 2.
In order to calculate the current fluctuations, 
we need to take into account that this distribution varies in time.
We denote the time-dependent distribution by
$\tilde{f}_{12}(\varepsilon,t)$.
The fluctuating current through conductor 1 or 2
causes electrostatic potential fluctuations $\delta \phi_{12}(t)$ 
in the reservoir, which enforce charge neutrality.
In Ref.\ \cite{b&b92}, the reservoir has a Fermi-Dirac
distribution
$\tilde{f}_{12}(\varepsilon,t)=
f_0[\varepsilon-e V_{12} - e \delta \phi_{12}(t)]$, with
$E_F+eV_{12}$ the average electrochemical potential in the reservoir.
As a result, it is found that the shot-noise power $P$
of the entire conductor is
given by \cite{b&b92}
\begin{equation}
R^2 P = R_1^2 P_1 + R_2^2 P_2 \: .
\label{s5e5}
\end{equation}
In other words, the voltage fluctuations add.
The noise powers $P_1$ and $P_2$ of the two segments
depend solely on the 
time-averaged distribution \cite{but90},
\begin{mathletters}
\label{s5e6}
\begin{eqnarray}
P_1 &=& 2 G_1 \!\! \int \!\! d\varepsilon \, [ f_L (1-f_L) + f_{12} ( 1-f_{12})]
+ 2 S_1 \!\! \int \!\! d\varepsilon \, (f_L -f_{12})^2 \: ,
\label{s5e6a}\\
P_2 &=& 2 G_2 \!\! \int \!\! d\varepsilon \, [ f_{12} ( 1-f_{12}) + f_R (1-f_R)]
+ 2 S_2 \!\! \int \!\! d\varepsilon \,(f_{12} -f_R)^2 \: .
\label{s5e6b}
\end{eqnarray}
\end{mathletters}%
Here, $S_m$ is defined as 
\begin{equation}
S_m = G_0  \mbox{Tr} \, {\sf t}_m^{\vphantom{\dagger}} {\sf t}_m^\dagger
({\sf 1} - {\sf t}_m^{\vphantom{\dagger}} {\sf t}_m^\dagger) = 
G_0 \sum_{n=1}^N T_n^{(m)} (1 - T_n^{(m)}) \: .
\label{s5e7}
\end{equation}
For example, for a single tunnel barrier we have $S_m = G_m$, 
whereas for a diffusive conductor $S_m=\frac{1}{3}G_m$.
The analysis of Ref.\ \cite{b&b92} is easily generalized to
arbitrary distribution $f_{12}$.
Then, we have 
$\tilde{f}_{12}(\varepsilon,t)=
f_{12}[\varepsilon - e \delta \phi_{12}(t)]$.
It follows that Eqs.\ (\ref{s5e5}) and (\ref{s5e6})
remain valid, but $f_{12}(\varepsilon)$ may be different.
Let us determine the shot noise for the three types of scattering.

(a) {\em Quasi-elastic scattering.} 
Here, it is not just the total current 
which must be conserved, but the current in each energy range.
This requires
\begin{equation}
f_{12}(\varepsilon) = 
\frac{ G_1 f_L(\varepsilon) + G_2 f_R(\varepsilon)}{G_1 + G_2} \: .
\label{s5e8}
\end{equation}
We note that Eq.\ (\ref{s5e8}) implies the validity of Eq.\ (\ref{s5e4}).
Substitution of Eq.\ (\ref{s5e8}) into Eqs.\ (\ref{s5e5}) and (\ref{s5e6})
yields at zero temperature the result
\begin{equation}
P= P_{\rm Poisson} \left( R_1^4 S_1 + R_2^4 S_2 + R_1 R_2^2 + R_1^2 R_2
\right) R^{-3}
\: .
\label{s5e9}
\end{equation}
For a double-barrier junction 
in the limit $\Gamma_1, \Gamma_2 \ll 1$,
Eqs.\ (\ref{e19}) and (\ref{s5e9}) give the same result, 
demonstrating that dephasing between the barriers does not influence the
shot noise.
This is in contrast to the result of Ref.\ \cite{dav95}, where dephasing is
modeled by adding random phases to the wave function.
For the diffusive wire Eq.\ (\ref{s5e9}) implies
$P= \case{1}{3} P_{\rm Poisson}$,
{\em independent} of the ratio between $R_1$ and $R_2$.
Breaking phase coherence, but retaining the nonequilibrium electron
distribution leaves the shot noise unaltered. 
The reservoir model for phase-breaking scattering
is therefore consistent with the results of the Boltzmann-Langevin approach.

(b) {\em Electron heating.} 
We model electron-electron
scattering, where energy can be exchanged between the electrons, 
at constant total energy.
We assume that the exchange of energies establishes a 
Fermi-Dirac 
distribution $f_{12}(\varepsilon)$ 
at an  electrochemical potential $E_F + e V_{12}$
and an elevated temperature $T_{12}$.
 From current conservation, Eq.\ (\ref{s5e3}), it follows that
\begin{equation}
V_{12} = \frac{R_2}{R} \, V \: .
\label{s5e10}
\end{equation}
Conservation of the energy of the electron system requires that
$T_{12}$
is such that no energy is
absorbed or emitted by the reservoir. 
The energy current $J_m$ through conductor $m$ is given by
\begin{mathletters}
\label{s5e11} 
\begin{eqnarray}
J_1 &=& (G_1/e^2) \! \int \!\! d\varepsilon \, 
\varepsilon \, [ f_L(\varepsilon) - f_{12}(\varepsilon) ]
\: ,
\\
J_2 &=& (G_2/e^2) \! \int \!\! d\varepsilon \, 
\varepsilon \, [ f_{12}(\varepsilon) - f_R(\varepsilon) ]
\: .
\end{eqnarray}
\end{mathletters}%
Since $f_{12}$ is a Fermi-Dirac distribution, Eq.\ (\ref{s5e11}) equals
\begin{mathletters}
\label{s5e12} 
\begin{eqnarray}
J_1 &=& Q_1 + \mu_1 I/e = K_1 (T_0- T_{12}) + \mu_1 I/e
\: , \\
J_2 &=& Q_2 + \mu_2 I/e = K_2 (T_{12} - T_0) + \mu_2 I/e
\: ,
\end{eqnarray}
\end{mathletters}%
where $\mu_1 \equiv E_F + \frac{1}{2}e(V + V_{12})$
and $\mu_2 \equiv E_F + \frac{1}{2}eV_{12}$.
The energy current $J_m$ is thus the sum of the heat current $Q_m$ and of
the particle current $I/e$ times the average energy $\mu_m$ 
of each electron.
The heat current $Q_m$ equals the difference in temperature times
the thermal conductance $K_m = G_m {\cal L}_0 T_m$, with
$T_m \equiv (T_0 + T_{12})/2$ and the Lorentz number 
${\cal L}_0 \equiv (k_B/e)^2 \pi^2/3$. 
There are no thermo-electric contributions
in Eqs.\ (\ref{s5e1}) and
(\ref{s5e12}),
because of the 
assumption of energy independent transmission eigenvalues \cite{hou92}.
 From the requirement of
energy conservation, $J_1=J_2$,
we calculate the electron temperature in the intermediate reservoir:
\begin{equation}
T_{12}^2 = T_0^2 + \frac{V^2}{{\cal L}_0} \, 
\frac{R_1 R_2}{R^2}
\: .
\label{s5e13}
\end{equation}
At zero temperature in the left and right reservoir and for $R_1=R_2$
we have $k_B T_{12}=(\sqrt{3}/2\pi) e |V| \simeq 0.28 e |V|$.
For the shot noise at $T_0=0$,
we thus obtain using Eqs.\ (\ref{s5e5}) and (\ref{s5e6}),
\begin{eqnarray}
P &=& 2 \frac{k_B T_{12}}{R} 
\nonumber\\
&+&2 \frac{S_1 R_1^2}{R^2}
\left\{ e(V-V_{12}) + k_B T_{12} 
[ 2 \ln( 1 + e^{e(V_{12}-V)/ k_B T_{12}} )
- 1] \right\}
\nonumber \\ &+&
2 \frac{S_2 R_2^2}{R^2}
\left\{ eV_{12} + k_B T_{12} [ 2 \ln( 1 + e^{-eV_{12}/ k_B T_{12}} )
- 1] \right\}   \: .
\label{s5e14}
\end{eqnarray}
The shot noise for two equal ($R_1=R_2$)
diffusive conductors,
\begin{equation}
P= P_{\rm Poisson} \frac{1}{\pi\sqrt{3}} \left[ 1+ \ln 2 + 
\ln \cosh(\pi/2\sqrt{3}) \right]
\simeq 0.38 \, P_{\rm Poisson}
\: ,
\label{s5e15}
\end{equation}
is slightly above the one-third suppression.
This shows that the current becomes less correlated due to the
electron-electron scattering.

(c) {\em Inelastic scattering.} 
This is the model of Ref.\ \cite{b&b92}.
The distribution function of the intermediate reservoir is 
the Fermi-Dirac distribution at the lattice
temperature $T_0$, with an
electrochemical potential $\mu_{12}\equiv E_F + e V_{12}$,
where $V_{12}$ is given by Eq.\ (\ref{s5e10}).
This reservoir absorbs energy, in contrast to cases (a) and (b).
The zero-temperature shot-noise power follows from
Eqs.\ (\ref{s5e5})--(\ref{s5e6}) \cite{b&b92}
\begin{equation}
P = P_{\rm Poisson} \frac{  R_1^3 S_1 + R_2^3 S_2}{R^2} \: .
\label{s5e16}
\end{equation}
For the diffusive case, with $R_1=R_2$, 
one has $P= \frac{1}{6} P_{\rm Poisson}$.
The inelastic scattering gives an additional suppression \cite{b&b92}.

For a double-barrier system it is plausible to model the
additional scattering by a single reservoir
between the barriers.
In a diffusive conductor, however, these scattering processes occur 
throughout the system. 
It is therefore more realistic
to divide the conductor
into $M$ segments, connected by reservoirs.
Equation (\ref{s5e5}) becomes
\begin{equation}
R^2 P = \sum_{m=1}^M R_m^2 P_m \: ,
\label{s5e17}
\end{equation}
where the noise power $P_m$ of segment $m$ is calculated
analogous to Eq.\ (\ref{s5e6}).
We take the 
continuum limit $M \rightarrow \infty$.
The electron distribution at position $x$ is denoted by
$f(\varepsilon,x)$.
At the ends of the conductor $f(\varepsilon,0)=f_L(\varepsilon)$ and
$f(\varepsilon,L)=f_R(\varepsilon)$, {\em i.e.}\
the electrons are
Fermi-Dirac distributed at temperature $T_0$
and with electrochemical potential $\mu(0) = E_F + e V$ and
$\mu(L)=E_F$, respectively.
The value of $f(\varepsilon,x)$ inside the conductor depends on the type of
scattering, (a), (b), or (c), and is determined below.

In the expression for $P_m$
only the first term of Eq.\ (\ref{s5e6a}) remains.
It follows from Eq.\ (\ref{s5e17}) that
the noise power is given by 
\begin{equation}
P= \frac{4}{R^2} \int \limits_0^L \! d x \, \frac{\rho(x)}{A} 
\int \!\! d \varepsilon 
f(\varepsilon,x) \, [1 - f(\varepsilon,x)] \: ,
\label{s5e18}
\end{equation}
where $\rho(x)$ is the resistivity at position $x$.
The total resistance is given by
\begin{equation}
R= \frac{1}{A} \int \limits_0^L \! d x \, \rho(x) \: .
\label{s5e19}
\end{equation}
For a constant resistivity $\rho$ we find from Eq.\ (\ref{s5e18}) 
\begin{equation}
P= \frac{4}{R L} \int \limits_0^L \! d x \! \int \!\! d \varepsilon 
\, f(\varepsilon,x) [1 - f(\varepsilon,x)] \: .
\label{s5e20}
\end{equation}
This formula 
has been derived by Nagaev from the Boltzmann-Langevin equation
for isotropic impurity scattering in the diffusive limit \cite{nag92}.
Our semiclassical calculation in the previous Sections is worked out in
terms of transmission probabilities rather than in terms of the electron
distribution function. However, one can easily convince oneself that in the
diffusive limit and at zero temperature,
Eqs.\ (\ref{3e7}) and (\ref{s5e20}) are equivalent.
The present derivation shows that 
the quantum mechanical expression for the noise
with phase-breaking reservoirs leads to the same
result as the
semiclassical approach.
We evaluate Eq.\ (\ref{s5e20}) for the three types of scattering.

(a) {\em Quasi-elastic scattering.} 
This calculation has previously been performed 
by Nagaev \cite{nag92} and is similar to Section \ref{s3}.
Current conservation and the absence of inelastic scattering
requires
\begin{equation}
\frac{\partial^2}{\partial x^2} f(\varepsilon,x) = 0 \: .
\label{s5e21}
\end{equation}
The solution is
\begin{equation}
f(\varepsilon,x) = 
\frac{L-x}{L} f(\varepsilon,0) + \frac{x}{L} f(\varepsilon,L)
\: ,
\label{s5e22}
\end{equation}
The electron distribution at $x=L/2$ is plotted in the left inset of
Fig.\ \ref{s5f2}.
Substitution of Eq.\ (\ref{s5e22})
into Eq.\ (\ref{s5e20}) yields \cite{nag92}
\begin{equation}
P=\frac{2}{3R} \left[ 4 k_B T_0 + e V \coth(eV/2 k_B T_0) \right] \: .
\label{s5e23}
\end{equation}
At zero temperature the shot noise is one-third of the Poisson noise.
The temperature dependence of $P$ is given in Fig.\ \ref{s5f2}.

(b) {\em Electron heating.}
This calculation is due to
Martinis and Devoret \cite{dev}. Similar derivations on the basis of the
Boltzmann-Langevin equation have been given by Nagaev \cite{nag95} and by
Kozub and Rudin \cite{koz95}.
The electron distribution function 
is a Fermi-Dirac distribution at an elevated
temperature $T_e(x)$,
\begin{equation}
f(\varepsilon,x)=
\left\{ 1+ \exp \left[ \frac{\varepsilon-\mu(x)}{k_B T_e(x)} \right]
\right\}^{-1} \: .
\label{s5e24}
\end{equation}
The current density $j(x)$ at $x$ is 
\begin{equation}
j(x)=- e D {\cal D}(E_F) \frac{\partial}{\partial x} 
\int \!\! d \varepsilon \, f(\varepsilon,x)  \: ,
\label{s5e25}
\end{equation}
where $D$ is the diffusion constant and ${\cal D}$ is the
density of states. 
We neglect the energy dependence of $D$ and ${\cal D}$.
The resistivity $\rho$ is given by the
Einstein relation, $\rho^{-1}=e^2  D {\cal D}(E_F)$.
Current conservation yields
\begin{equation}
\frac{\partial j(x)}{\partial x} =0 \: ,
\label{s5e26}
\end{equation}
which implies for the electrochemical potential 
\begin{equation}
\mu(x) = E_F + \frac{L-x}{L} e V \: .
\label{s5e27}
\end{equation}
The energy-current density $j_\varepsilon(x)$ is determined according to
\begin{mathletters}
\label{s5e28}
\begin{eqnarray}
j_\varepsilon(x)&=&- D  {\cal D}(E_F)
\frac{\partial}{\partial x} \int \!\! d \varepsilon \,
\varepsilon f(\varepsilon,x) 
=\mu(x) j(x)/e + j_Q(x) \: ,
\label{s5e28a} \\
j_Q(x) &\equiv&  - \kappa(x) \frac{\partial T_e(x)}{\partial x}
\: .
\label{s5e28b}
\end{eqnarray}
\end{mathletters}%
The heat-current density $j_Q(x)$ equals the temperature gradient times
the heat conductivity $\kappa(x) = T_e(x) {\cal L}_0 / \rho$.
Because of energy conservation
the divergence of the energy-current density must be zero,
\begin{equation}
\frac{\partial j_\varepsilon(x)}{\partial x} = 0
\: .
\label{s5e29}
\end{equation}
Combining Eqs.\ (\ref{s5e28}) and (\ref{s5e29}), we obtain the following
differential equation for the temperature
\begin{equation}
\frac{\partial^2}{\partial x^2} [T_e(x)^2] = 
- \frac{2}{{\cal L}_0}  \left( \frac{V}{L} \right)^2
\: .
\label{s5e30}
\end{equation}
Taking into account the boundary conditions the solution is
\begin{equation}
T_e(x)=\sqrt{ T_0^2 + (x/L)[1-(x/L)] \, V^2 /{\cal L}_0 }
\: .
\label{s5e31}
\end{equation}
In the middle of the wire the electron temperature takes its
maximum value. For zero lattice temperature ($T_0=0$)
one has $k_B T_e(L/2)= (\sqrt{3} / 2 \pi) e |V| \simeq 0.28 e |V|$.
The electron distribution at $x=L/2$ is depicted in the left inset
and
the electron temperature profile (\ref{s5e31}) is plotted in
the right inset of Fig.\ \ref{s5f2}.

Equations (\ref{s5e20}), (\ref{s5e24}), and (\ref{s5e31})
yields
for the noise power the result
\begin{eqnarray}
P&=& \frac{4}{R L} \int \limits_0^L \! dx \, k_B T_e(x)
\nonumber \\
&=& \frac{2 k_B T_0}{R} + 2 e I \left[ \frac{2\pi}{\sqrt{3}} 
\left( \frac{k_B T_0}{e V} \right)^2 + \frac{\sqrt{3}}{2\pi} \right]
\arctan \left( \frac{\sqrt{3}}{2\pi} \, \frac{e V}{k_B T_0} \right)
\: .
\label{s5e33}
\end{eqnarray}
Equation (\ref{s5e33}) is plotted in Fig.\ \ref{s5f2}.
For the limit $eV \gg k_B T_0$ one finds \cite{ste95}
\begin{equation}
P= \case{1}{4} \sqrt{ 3 } \, P_{\rm Poisson} \simeq  0.43 \, P_{\rm Poisson}
\: .
\label{s5e34}
\end{equation}
Due to the electron-electron scattering the shot noise is increased.
The exchange of energies among the electrons makes the current less
correlated.
The suppression factor of $\frac{1}{4}\sqrt{3}$ is close to the
value observed in an experiment on silver wires
by Steinbach, Martinis, and Devoret \cite{ste95}.

(c) {\em Inelastic scattering.} 
The electron distribution function is given by
\begin{equation}
f(\varepsilon,x)=
\left\{ 1+ \exp \left[ \frac{\varepsilon-\mu(x)}{k_B T_0} \right]
\right\}^{-1} \: ,
\label{s5e35}
\end{equation}
with $\mu(x)$ according to Eq.\ (\ref{s5e27}).
For the noise power we obtain from Eqs.\ (\ref{s5e20}) and (\ref{s5e35})
\begin{equation}
P= \frac{4 k_B T_0}{R}\: ,
\label{s5e36}
\end{equation}
which is equal to the Johnson-Nyquist noise 
for arbitrary $V$ (see Fig.\ \ref{s5f2}). 
The shot noise is thus
completely suppressed by the inelastic scattering
\cite{b&b92,shi92,nag95,koz95,lan93,liu94}.

These calculations assume a constant cross-section and resistivity
of the conductor. 
One might wonder, whether variations in
cross-section and resistivity, which will certainly appear in experiments,
change the one-third suppression for the case of elastic scattering and
the $\case{1}{4}\sqrt{3}$ suppression for the case of electron-heating.
In Appendix \ref{a4}, it is demonstrated how this can be calculated 
on the basis of Eq.\ (\ref{s5e18}).
It is found that the results 
[Eqs.\ (\ref{s5e23}), (\ref{s5e33}), and (\ref{s5e36})] are independent of
smooth variations in cross-section and resistivity.
We thus conclude, that both the one-third suppression
as well as the $\case{1}{4}\sqrt{3}$ suppression are in principle
observable in {\em any} diffusive conductor.

\section{Conclusions and discussion}
\label{s6}

We have derived a general formula
for the shot noise within the framework of the semiclassical 
Boltzmann-Langevin equation.
We have applied this to the case of a
disordered conductor,
where we have calculated how the shot noise crosses over from
complete suppression in the ballistic limit to one-third of the Poisson noise
in the diffusive limit. 
Furthermore, we have applied our formula 
to the shot noise in a conductor consisting of a 
sequence of tunnel barriers. 
Finally, we have considered a disordered conductor in series with a tunnel
barrier.
For all these systems, we have
obtained a sub-Poissonian shot-noise power, in complete agreement with
quantum mechanical calculations in the literature.
This establishes that phase coherence is not required for the
occurrence of suppressed shot noise in mesoscopic conductors.
Moreover, it has been shown that for diffusive conductors the one-third
suppression occurs quite generally. This phenomenon depends neither 
on the dimensionality of the conductor, nor on the microscopic
details of the scattering potential.

We have modeled quasi-elastic scattering (which breaks phase coherence),
electron heating (due to electron-electron scattering),
and inelastic scattering (due to, {\em e.g.}, electron-phonon scattering)
by putting charge-conserving reservoirs between phase-coherent
segments of the conductor.
If the scattering occurs throughout the whole length of the conductor, we
end up with the same formula for the noise as can be obtained
directly from the Boltzmann-Langevin approach \cite{nag95,koz95}.
In the case of electron heating, the shot noise is $\case{1}{4}\sqrt{3}$
of the Poisson noise, which is slightly
above $\case{1}{3} P_{\rm Poisson}$ for the fully elastic case. 
The experiments of Refs.\ \cite{lie94} and \cite{ste95}
are likely in this
electron-heating regime.
We have demonstrated that both the one-third suppression and 
the $\case{1}{4} \sqrt{3}$ suppression are insensitive to 
the geometry of the conductor, 
as long as the transport is in the diffusive regime.
For future work,
it might be worthwhile to take the effects of electron heating and 
inelastic scattering into account through the scattering terms in the 
Boltzmann-Langevin equation, as has been done in Refs.\ \cite{nag95,koz95},
in order to calculate the crossover between the
different regimes.

In both the quantum mechanical and semiclassical theories
the electrons are treated as noninteracting particles.
Some aspects of the electron-electron interaction are taken into account
by the conditions on the reservoirs in Section \ref{s5}, 
where fluctuations in the
electrostatic potential enforce charge-neutrality.
We have shown that these fluctuations
suppress the noise only in the presence 
of inelastic scattering.
Coulomb repulsion is known to have a strong effect on the noise in
confined geometries with a small capacitance \cite{her92,kor92}.
This is relevant for the double-barrier case treated in Section
\ref{s4}.
Theories which take the Coulomb blockade into account \cite{her92,kor92} 
predict a shot-noise suppression which is periodic in the
applied voltage.
This effect has recently been observed for a
nanoparticle between a surface and the tip of a scanning tunneling
microscope \cite{bir95}.
In open conductors we would expect
these interaction effects to be less important \cite{but93}.

\acknowledgements

We thank M. H. Devoret and R. Landauer for valuable discussions.
This research has been
supported by the ``Ne\-der\-land\-se
or\-ga\-ni\-sa\-tie voor We\-ten\-schap\-pe\-lijk On\-der\-zoek'' (NWO)
and by the ``Stich\-ting voor Fun\-da\-men\-teel On\-der\-zoek der
Ma\-te\-rie'' (FOM).

\appendix

\section{Thermal noise}
\label{a1}

In this Appendix it is shown how thermal fluctuations can be incorporated 
in the theory.
These are ignored in Sections \ref{s3} and \ref{s4}
where zero temperature is considered.
At non-zero temperatures we need to take into account
the time-dependent fluctuations in the
occupation of the states in the reservoirs.
The formal solution of the Boltzmann-Langevin
equation (\ref{e5}) can be written as
\begin{eqnarray}
\delta f ({\bf r}, {\bf k}, t)&=&
\int \limits_{-\infty}^{t} \! \! 
dt' \int \limits_{\cal V} d{\bf r}' \int \! d{\bf k}'
{\cal G}( {\bf r}, {\bf k}; {\bf r}', {\bf k}'; t-t') \,
j({\bf r}', {\bf k}', t') 
\nonumber \\
&+& 
\int \limits_{-\infty}^{t} \! \! dt'
\int \limits_{S_L}  \!\! d{\bf y}' \! 
\int \limits_{k_x>0} \!\! d {\bf k}' \, v_x \,
{\cal G}( {\bf r}, {\bf k}; {\bf r}', {\bf k}'; t-t') \,
\delta f({\bf r}', {\bf k}',t')
\nonumber \\
&+& 
\int \limits_{-\infty}^{t} \! \! dt'
\int \limits_{S_R}  \!\! d{\bf y}' \! 
\int \limits_{k_x<0} \!\! d {\bf k}' \, |v_x| \,
{\cal G}( {\bf r}, {\bf k}; {\bf r}', {\bf k}'; t-t') \,
\delta f({\bf r}', {\bf k}',t')  \: ,
\label{ae1}
\end{eqnarray}
where $\cal V$ denotes the scattering region of the conductor.
The second and third term describe the time-dependent
fluctuations of  states 
originating from the reservoir which are ignored in Eq.\ (\ref{e12}).
The correlation function of the 
incoming fluctuations which have not yet reached the scattering region
[{\em i.e.}\ for the left lead  $x,x' \leq x_L$, $k_x, k_x' > 0$ and for 
the right lead $x,x' \geq x_R$, $k_x, k_x' < 0$]
follows from Eqs.\ (\ref{ee1}) and (\ref{ee2}):
\begin{equation}
\langle \delta f({\bf r}, {\bf k}, t)  \, \delta f({\bf r}', {\bf k}', t') 
\rangle = (2 \pi)^d \,
\delta[ {\bf r} - {\bf r}' - {\bf v}(t-t') ] \, \delta( {\bf k} - {\bf k'})
f_{L,R}(\varepsilon) [ 1 - f_{L,R}(\varepsilon)]
\: .
\label{ae2}
\end{equation}
The derivation of the noise power proceeds similar to the derivation of 
Eq.\ (\ref{e12}). 
Substitution of Eq.\ (\ref{ae1}) into Eqs.\ (\ref{e7}) and 
(\ref{e8})  and using both the correlation functions (\ref{e6}) and
(\ref{ae2}), yields
\begin{eqnarray}
P&=&\frac{2 e^2}{ (2 \pi)^d} \left( \int \limits_{\cal V} \!  d{\bf r} \! 
\int \! d {\bf k} \! \int  \! d {\bf k}' \,
T({\bf r},{\bf k}) T({\bf r},{\bf k}')
J({\bf r},{\bf k},{\bf k}') \right.
\nonumber \\
&& + \int \limits_{S_L} \!\! d{\bf y} \! 
\int \limits_{k_x>0}  \!\! d {\bf k} \, v_x  \, T({\bf r},{\bf k})^2 
f_L(\varepsilon) [ 1 - f_L(\varepsilon)] 
\nonumber \\
&& \left . + \int \limits_{S_R}  \!\! d{\bf y} \! 
\int \limits_{k_x<0}  \!\! d {\bf k} \, |v_x| \, [1-T({\bf r},{\bf k})]^2 
f_R(\varepsilon) [ 1 - f_R(\varepsilon)] \right) 
\: .
\label{ae3}
\end{eqnarray}

Let us apply Eq.\ (\ref{ae3}) to the case of impurity scattering,
 treated in Section \ref{s3} for zero temperature.
By changing variables according to Eq.\ (\ref{3ee1}) and
by substitution of Eqs.\ (\ref{ee5}) and (\ref{3e4}), we obtain
\begin{eqnarray}
P &=& 2 e^2 A \int \limits_0^L \! \! dx \int \!\! d\varepsilon \, 
{\cal D}(\varepsilon)
\int \frac{d {\bf \hat{n}}}{s_d} 
\int \frac{d {\bf \hat{n}}'}{s_d} \,
W_{{\bf \hat{n}}{\bf \hat{n}}'} \,
[ T(x,{\bf \hat{n}}) - T(x,{\bf \hat{n}}')]^2
\nonumber \\ &&
\mbox{} \times \left\{ \vphantom{\int}  f_L(\varepsilon)[1-f_L(\varepsilon)]
\,  [1 - T(x,-{\bf \hat{n}})]  + f_R(\varepsilon)[1-f_R(\varepsilon)] \,
T(x,-{\bf \hat{n}}) \right.
\nonumber \\ &&
\hphantom{\mbox{} \times} + \left. [f_L(\varepsilon)-f_R(\varepsilon)]^2 
T(x,-{\bf \hat{n}})[1-T(x,-{\bf \hat{n}}')] \vphantom{\int} \right\}
\nonumber \\ &+&
2 e^2 A \int \!\! d\varepsilon \, {\cal D}(\varepsilon) 
f_L(\varepsilon)[1-f_L(\varepsilon)] 
\int \frac{d {\bf \hat{n}}}{s_d} \, v \, n_x \, T^2(0,{\bf \hat{n}})
\nonumber \\ &-&
2 e^2 A \int \!\! d\varepsilon \, {\cal D}(\varepsilon) 
f_R(\varepsilon)[1-f_R(\varepsilon)] 
\int \frac{d {\bf \hat{n}}}{s_d} \, v \, n_x \, [1-T(L,{\bf \hat{n}})]^2
\: ,
\label{ae4}
\end{eqnarray}
where we have used Eq.\ (\ref{3e1}) and 
$W_{{\bf \hat{n}}{\bf \hat{n}}'} =W_{{\bf \hat{n}}'{\bf \hat{n}}}$.
Equation (\ref{ae4}) 
can be simplified by means of the relations
\begin{equation}
\int \limits_0^L \! \! dx \int \frac{d {\bf \hat{n}}}{s_d} 
\int \frac{d {\bf \hat{n}}'}{s_d} \,
W_{{\bf \hat{n}}{\bf \hat{n}}'} \,
[ T(x,{\bf \hat{n}}) - T(x,{\bf \hat{n}}')]^2
= v_F \int \frac{d {\bf \hat{n}}}{s_d} \, n_x \,
[ T^2(L,{\bf \hat{n}}) - T^2(0,{\bf \hat{n}})] \: ,
\label{ae5}
\end{equation}
\begin{eqnarray}
&&\int \limits_0^L \! \! dx \int \frac{d {\bf \hat{n}}}{s_d} 
\int \frac{d {\bf \hat{n}}'}{s_d} \,
W_{{\bf \hat{n}}{\bf \hat{n}}'} \,
[ T(x,{\bf \hat{n}}) - T(x,{\bf \hat{n}}')]^2 T(x,-{\bf \hat{n}})
\nonumber \\
&&
\hspace{1cm} = v_F \int \frac{d {\bf \hat{n}}}{s_d} \, n_x \,
[ T^2(L,{\bf \hat{n}}) T(L,-{\bf \hat{n}})  - 
T^2(0,{\bf \hat{n}}) T(0,-{\bf \hat{n}})] \: ,
\label{ae6}
\end{eqnarray}
which can be derived from Eq.\ (\ref{3e1}).
For the distribution function we apply the identity
$f_0(1-f_0)=-k_B T_0 \, \partial f_0/\partial\varepsilon$ and
define
\begin{equation}
F(V,T_0)\equiv \int d\varepsilon  
[f_L(\varepsilon)-f_R(\varepsilon)]^2 = e |V| \coth 
\left(\frac{e |V|}{2 k_B T_0} \right) - 2 k_B T_0 \: .
\label{ae7}
\end{equation}
Collecting results, we find for the noise power the expression
\begin{eqnarray}
P &=& \frac{2 F(V,T_0) G_0 N}{v_F s_d v_{d-1}}
\int \limits_0^L \! \! dx
\int \! \! d {\bf \hat{n}} 
\int \! \! d {\bf \hat{n}}' \,
W_{{\bf \hat{n}}{\bf \hat{n}}'} 
\, [ T(x,{\bf \hat{n}}) - T(x,{\bf \hat{n}}')]^2  
T(x,-{\bf \hat{n}})[1-T(x,-{\bf \hat{n}}')]
\nonumber \\ &+&
4 k_B T_0 \, G_0 N  \int \frac{d {\bf \hat{n}}}{v_{d-1}} 
\, n_x T(L,{\bf \hat{n}}) \: .
\label{ae8}
\end{eqnarray}
At zero voltage, Eqs.\ (\ref{3e5}) and (\ref{ae8}) reduce to
the Johnson-Nyquist noise $P=4 k_B T_0 G$.
At zero temperature, Eq.\ (\ref{ae8}) reduces to Eq.\ (\ref{3e7}).
Applying Eq.\ (\ref{ae8}) to impurity scattering
for the case $d=1$ of Section
\ref{s3}, we obtain
\begin{equation}
P=\frac{2 G}{3} \left\{
e V \coth \left( \frac{e V}{2 k_B T_0} \right)
\left[1 - \frac{1}{(1+L/\tilde{\ell})^3} \right]
+ 2 k_B T_0
\left[2 + \frac{1}{(1+L/\tilde{\ell})^3} \right] \right \}
\: .
\label{ae9}
\end{equation}
The voltage dependence of the noise is plotted in Fig.\ \ref{af1}
for various values of $L/\tilde{\ell}$. 
The result for the diffusive limit is equal to Eq.\ (\ref{s5e23}).
Also depicted is the classical
result for a single high tunnel barrier ($\Gamma \ll 1$), 
\begin{equation}
P=2 e |I| \coth \left( \frac{e V}{2 k_B T_0} \right)
\: ,
\label{ae9a}
\end{equation}
which can be derived within our theory by combining the results of Section
\ref{s4} with the analysis of this Appendix.

\section{Noise at arbitrary cross section}
\label{a2}

Let us verify that 
the noise power does not depend on the 
location $x$ of the cross section at which the current is evaluated.
The fluctuating current through a cross section $S_x$
at coordinate $x$ is defined by
\begin{equation}
\delta I (t,x) \equiv \frac{e}{(2 \pi)^d} 
\int \limits_{S_x} \! \! d{\bf y} \int \! d {\bf k} \:
v_x \, \delta f ({\bf r}, {\bf k}, t) \: ,
\label{ae10}
\end{equation}
and leads to
\begin{equation}
P(x,x') \equiv 2 \int \limits_{-\infty}^{\infty} \! \!
dt \, \langle \delta I (t,x) \, \delta I (0,x') \rangle \: .
\label{ae11}
\end{equation}
We use the following relation
\begin{equation}
\int \limits_0^{\infty} \!\! dt 
\int \limits_{S_{x}} \!\! d{\bf y} 
\int \! d{\bf k} \, v_x \,
{\cal G}( {\bf r}, {\bf k}; {\bf r}_0, {\bf k}_0;t) =
T({\bf r}_0, {\bf k}_0) - \Theta(x_0-x) \: ,
\label{ae12}
\end{equation}
which follows from Eqs.\ (\ref{e10}) and (\ref{e11}).
Here, $\Theta(x)$ is the unit-step function.
Evaluating Eq.\ (\ref{ae11}) along the lines 
of Section \ref{s2}, we find
\begin{eqnarray}
P(x,x')&=&\frac{2 e^2}{ (2 \pi)^d} \int \!  d{\bf r}_0 \!  
\int \!  d {\bf k}_0 \!  \int \!  d {\bf k}'_0 \,
J({\bf r}_0,{\bf k}_0,{\bf k}_0')
\nonumber \\
&& \mbox{} \times 
[ T({\bf r}_0,{\bf k}_0)-\Theta(x_0-x) ]\, 
[T({\bf r}_0,{\bf k}'_0)-\Theta(x_0-x') ]\, 
 \: .
\label{ae13}
\end{eqnarray}
We use that the integral over ${\bf k}$ or over ${\bf k}'$
of $J({\bf r},{\bf k},{\bf k}')$ vanishes, Eq.\ (\ref{ae14}),
and find that $P(x,x')$ is independent of $x,x'$.

\section{Nonisotropic scattering}
\label{a3}

We wish to demonstrate that the occurrence of one-third suppressed shot noise
in the diffusive regime is independent of the angle-dependence
of the scattering rate. We write
$W_{{\bf \hat{n}}{\bf \hat{n}}'}=w({\bf \hat{n}}\cdot{\bf \hat{n}}') v_F$,
with arbitrary $w$.
In the diffusive limit, the transmission probability is given by
\begin{equation}
T(x,{\bf \hat{n}})=T(x) + t(n_x) \: ,
\label{ae21}
\end{equation}
where  $T(x)=x/L$ and $t(n_x)$ of order $\tilde{\ell} / L$, with
$\int d{\bf \hat{n}} \, t(n_x)=0$.
The conductance is given by the Drude result, Eq.\ (\ref{3e8a}), where the
normalized mean free path $\tilde{\ell}$ can be derived as follows:
Upon integration of Eq.\ (\ref{3e1a}) over $d {\bf \hat{n}} \, n_x$
and substitution of Eq.\ (\ref{ae21}), one
obtains
\begin{equation}
\int \frac{d {\bf \hat{n}}}{s_d} \,  n_x^2 \,
\frac{d T(x)}{d x} 
= \int \frac{d {\bf \hat{n}}}{s_d}
\int \frac{d {\bf \hat{n}}'}{s_d} \, n_x w({\bf \hat{n}}\cdot{\bf \hat{n}}')
\left[ t(n_x) - t(n'_x) \right] \: .
\label{ae21a}
\end{equation}
Comparison with Eq.\ (\ref{3e5}) yields
\begin{equation}
\tilde{\ell} = \frac{v_d}{v_{d-1}} \,
\left[ \int \frac{d {\bf \hat{n}}}{s_d} \, w(n_x) (1-n_x) \right]^{-1}
\: .
\label{ae22}
\end{equation}
 From Eq.\ (\ref{3e1a}) it also follows that 
\begin{eqnarray}
\int \frac{d {\bf \hat{n}}}{s_d}
\int \frac{d {\bf \hat{n}}'}{s_d} \, W_{{\bf \hat{n}}{\bf \hat{n}}'}
\,
[ T( x, {\bf \hat{n}} ) - T( x, {\bf \hat{n}}')]^2
&=& 
v_F  \frac{\partial}{\partial x}
\int \frac{d {\bf \hat{n}}}{s_d} \,
n_x  T^2( x, {\bf \hat{n}} )
\nonumber \\
&=& 
\frac{2 G v_F v_{d-1}}{G_0 N s_d} \frac{d T(x)}{dx} \: ,
\label{ae23}
\end{eqnarray}
where we have used Eqs.\ (\ref{3e5}) and (\ref{ae21}).
By substitution of Eq.\ (\ref{ae23}) into Eq.\ (\ref{3e7}) 
and neglecting terms of order $\tilde{\ell}/L$, we find
\begin{equation}
P=2 P_{\rm Poisson} \int \limits_0^L dx \, T(x)[1-T(x)] \frac{d T(x)}{dx} 
= \case{1}{3} P_{\rm Poisson}\: ,
\label{ae24}
\end{equation}
independent of $w$.

\section{The effect of variations in cross-section and resistivity}
\label{a4}

In Section \ref{s5}, we have calculated the shot noise in a diffusive
conductor for several types of scattering. It has been assumed that both
the area of the cross section $A$ and the resistivity $\rho$ are constant
along the conductor. Below, we briefly describe how the calculations are
modified by taking into account a non-constant, but smoothly varying
area $A(x)$ and resistivity $\rho(x)$.

Our starting point is Eq.\ (\ref{s5e18}). 
It is convenient to change variables from $x$ to $\eta$,
defined according to
\begin{equation}
\eta \equiv \frac{1}{R} \int \limits_{0}^{x} dx' \, \frac{A(x')}{\rho(x')} \: .
\label{ae41}
\end{equation}
In other words, $\eta$ is ratio between the resistance of the conductor
from 0 to $x$ and the total resistance.
Equation (\ref{s5e18}) thus becomes
\begin{equation}
P = \frac{4}{R} \int \limits_0^1 d\eta \int d \varepsilon 
f(\varepsilon,\eta) [ 1- f(\varepsilon,\eta) ] \: .
\label{ae42}
\end{equation}
It is now straightforward to repeat the calculation for the diffusive
conductor in Section \ref{s5}. It follows, that all the results 
[Eqs.\ (\ref{s5e23}), (\ref{s5e33}), and (\ref{s5e36})] remain unaltered.
Here, we will just illustrate how the calculation for the case of electron
heating is done.

Starting from Eq.\ (\ref{s5e24}) we find for the current at position $\eta$
\begin{equation}
I(\eta)=- \frac{1}{e R} \, \frac{\partial}{\partial \eta} \, \mu(\eta) \: .
\label{ae43}
\end{equation}
 From current conservation $\{ \, I(\eta)=I$ for all $\eta \in [0,1] \, \}$
it follows that the electro-chemical potential is
\begin{equation}
\mu(\eta)=E_{\rm F} + (1-\eta) e V \: .
\label{ae44}
\end{equation}
The energy current is given by
\begin{equation}
I_\varepsilon(\eta) = \frac{\mu(\eta) I}{e} - 
\frac{{\cal L}_0}{R} \, T_e(\eta) 
\frac{\partial}{\partial \eta} T_e(\eta) 
\: .
\label{ae45}
\end{equation}
Similarly to the derivation in Section \ref{s5}, we thus find
\begin{equation}
T_e(\eta) = \sqrt{ T_0^2 + \eta(1-\eta) V^2/{\cal L}_0 }\: ,
\label{ae46}
\end{equation}
from which it follows that the noise is given by Eq.\ (\ref{s5e33}),
as before.

\begin{figure}
\vspace{5cm}\centerline{\psfig{file=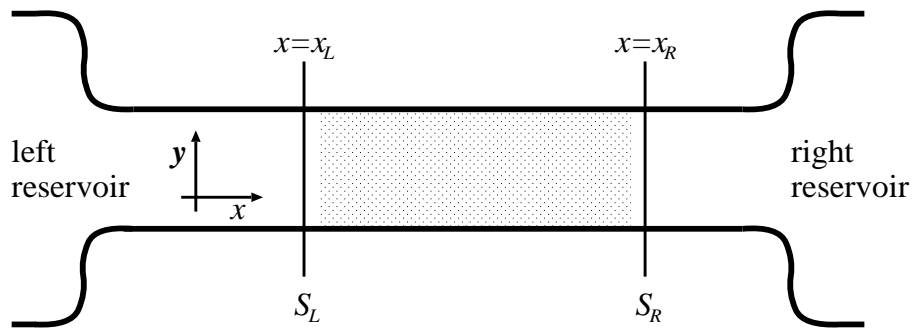,width=12cm}}\vfill
\caption{
The conductor consists of a scattering region (dotted)
connected by perfect
leads to two electron reservoirs. Cross sections
$S_L$ and $S_R$ in the left and right lead are indicated. 
}
\label{ff1}
\end{figure}

\newpage
\begin{figure}
\vspace{5cm}\centerline{\psfig{file=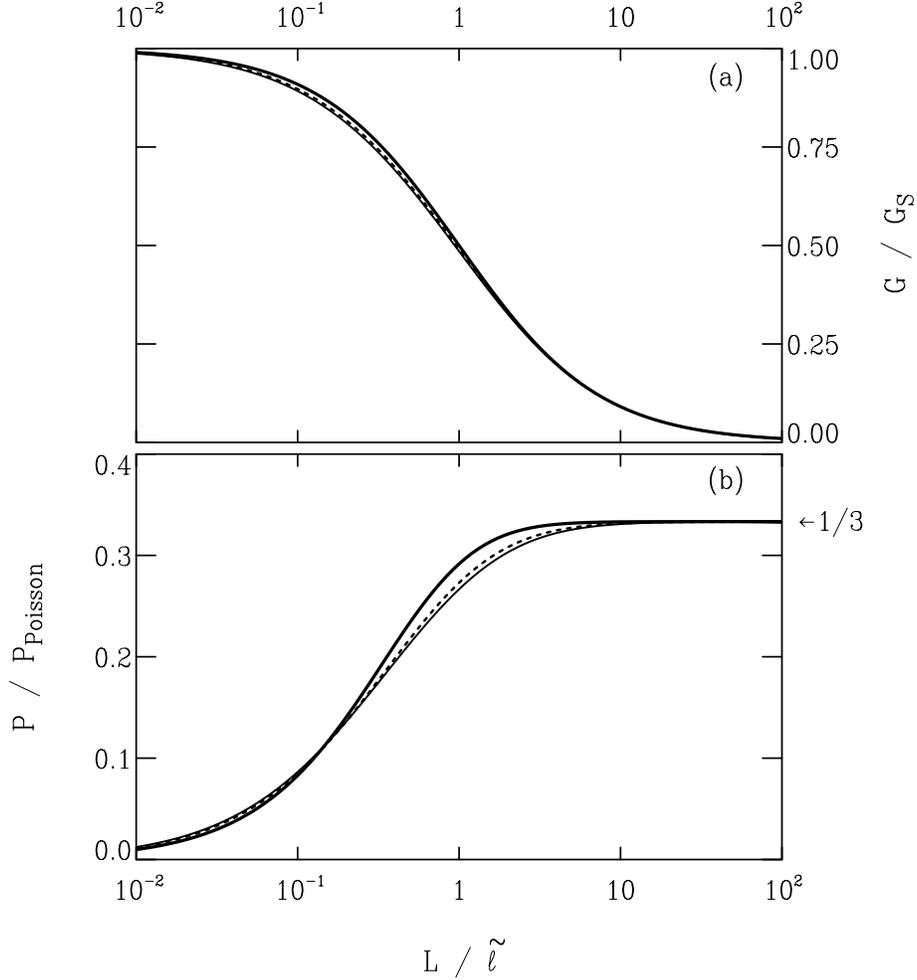,width=12cm}}\vfill
\caption{(a) The conductance 
(normalized by the Sharvin conductance $G_S=N G_0$)
and (b) the shot-noise power (in units of 
$P_{\rm Poisson}\equiv 2 e |\bar{I}|$), as a function of the ratio
$L/\tilde{\ell}$, computed from Eqs.\ (\ref{3e5}) and (\ref{3e7})
for isotropic impurity scattering. The curves correspond to
a three-dimensional (thin solid curve), two-dimensional (dashed curve), and a
one-dimensional conductor (thick solid curve).
The one-dimensional case is the analytical result from
Eqs.\ (\ref{3e11}) and (\ref{3e12}).
The two- and three-dimensional cases are numerical results.
}
\label{ff2}
\end{figure}

\newpage
\begin{figure}
\vspace{5cm}\centerline{\psfig{file=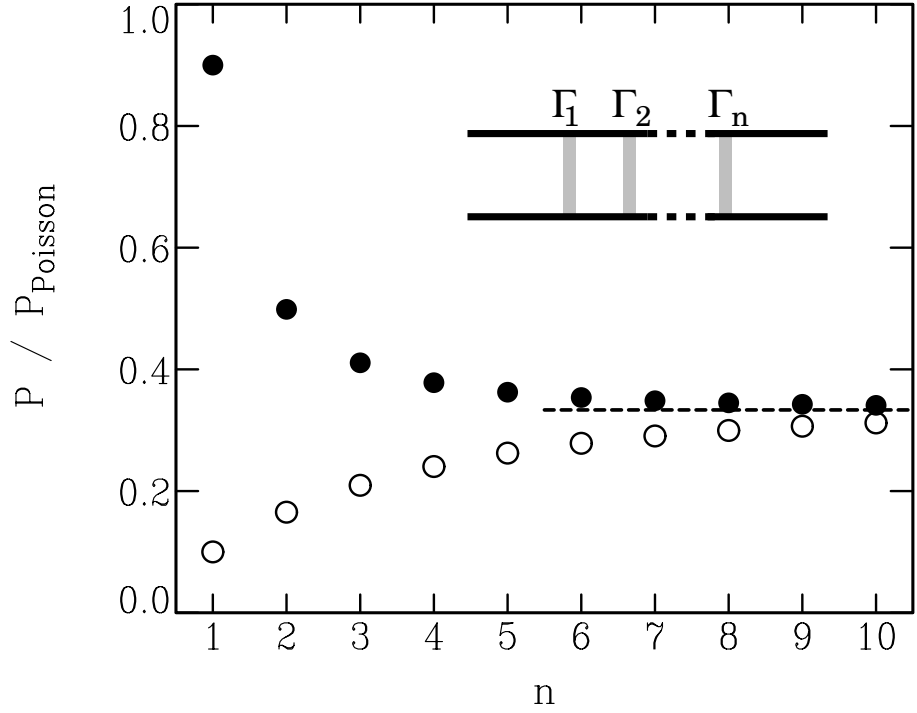,width=12cm}}\vfill
\caption{
The shot-noise power $P$ for
$n$ tunnel barriers in series with transmission probability
$\Gamma=0.1$ (dots) and $\Gamma=0.9$ (circles), computed from Eq.\
(\protect\ref{e22}). 
The dashed line is the large-$n$ limit $P=\case{1}{3} P_{{\rm Poisson}}$.
The inset shows schematically the geometry considered.}
\label{f1}
\end{figure}

\newpage
\begin{figure}
\vspace{5cm}\centerline{\psfig{file=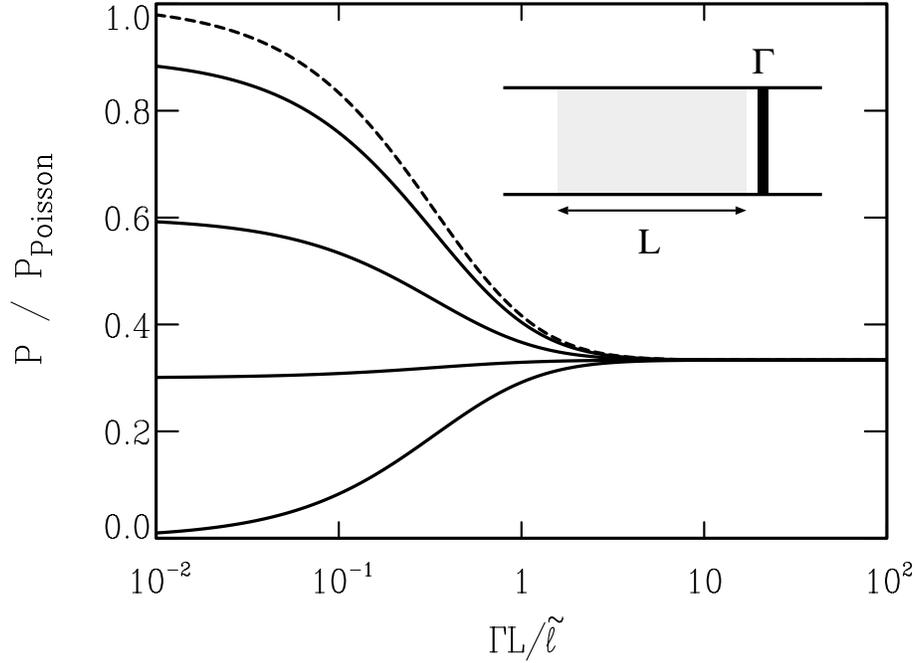,width=12cm}}\vfill
\caption{
The shot-noise power $P$ for a conductor
consisting of a disordered region in series with a planar tunnel
barrier (see inset)
as a function of its length $L$ (in units of $\tilde{\ell}/\Gamma$),
for barrier transparencies $\Gamma=$1, 0.7, 0.4, and 0.1
(bottom to top). The dashed line is the limiting curve 
for $\Gamma \ll 1$.
The curves are computed from Eq.\ (\ref{4e13}) for a model with a
one-dimensional density of states. The dimensionality dependence is
expected to be small, compare Fig.\ \ref{ff2}.
}
\label{ff3}
\end{figure}

\newpage
\begin{figure}
\vspace{5cm}\centerline{\psfig{file=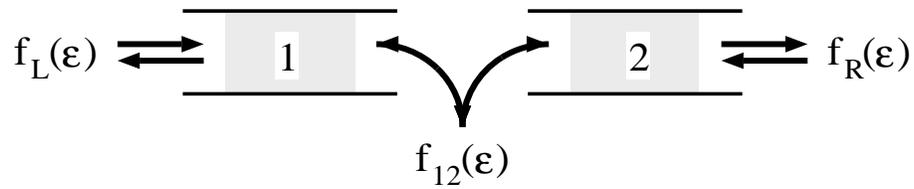,width=12cm}}\vfill
\caption{
Both ends of
the conductor are connected to an electron reservoir.
Additional scattering inside the conductor is modeled
by dividing it in two parts and connecting them
through another reservoir.
The electron distributions
in the left and the right reservoir, 
$f_L(\varepsilon)$ and $f_R(\varepsilon)$, are Fermi-Dirac
distributions. The distribution $f_{12}(\varepsilon)$ in the intermediate
reservoir depends on the type of scattering. 
}
\label{s5f1}
\end{figure}

\newpage
\begin{figure}
\vspace{5cm}\centerline{\psfig{file=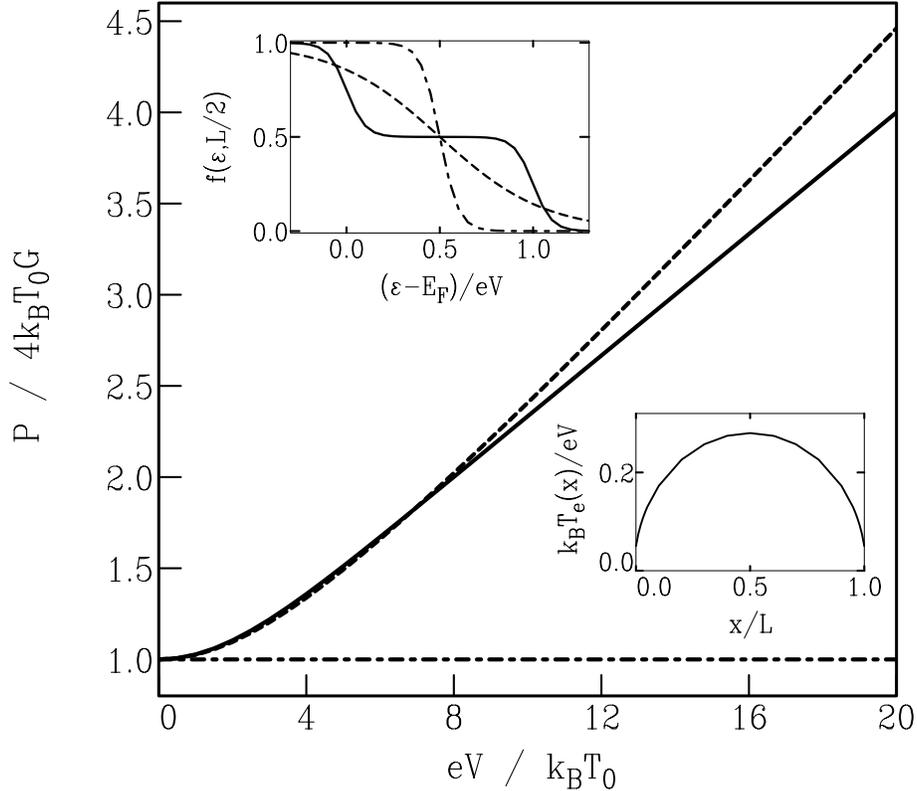,width=12cm}}\vfill
\caption{
The noise power $P$ (divided by the Johnson-Nyquist noise
$4 k_B T_0 G$) versus applied voltage $V$ for a
disordered wire for model (a) of quasi-elastic scattering (solid curve),
(b) of electron heating (dashed curve), 
(c) of inelastic scattering (dash-dotted curve),
according to Eqs.\ (\ref{s5e23}), (\ref{s5e33}), and (\ref{s5e36}), 
respectively.
The upper left inset gives the electron distribution
in the middle of the wire 
$f(\varepsilon,L/2)$ as a function of energy $\varepsilon$
for model (a), (b), and (c).
The lower right inset shows the temperature $T_e(x)$ 
as a function of the position
$x$ for model (b). For both insets, $k_B T_0= \case{1}{20} e |V|$.
}
\label{s5f2}
\end{figure}

\newpage
\begin{figure}
\vspace{5cm}\centerline{\psfig{file=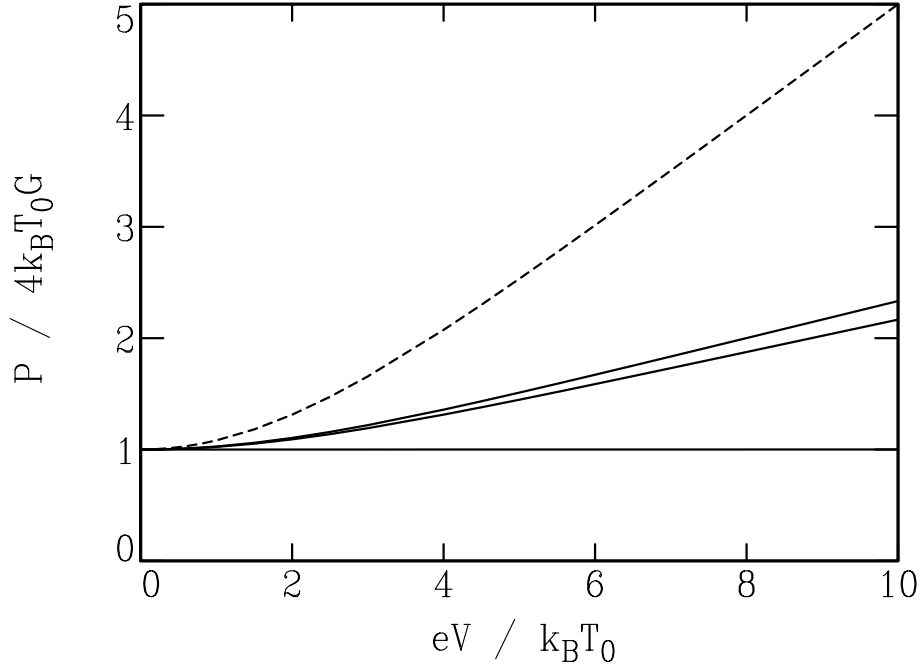,width=12cm}}\vfill
\caption{
The noise power $P$ (divided by the Johnson-Nyquist noise
$4 k_B T_0 G$) versus applied voltage $V$ for a
disordered wire (bottom to top) 
in the ballistic limit $L/\tilde{\ell} \rightarrow 0$,
the intermediate regime $L/\tilde{\ell}=1$, and in the
diffusive limit $L/\tilde{\ell} \rightarrow \infty$,
as given by Eq.\ (\ref{ae9}).
The dashed line is the noise in a tunnel barrier, according to
Eq.\ (\ref{ae9a}).
}
\label{af1}
\end{figure}

\end{document}